\documentclass{article}
\usepackage{amssymb}
\usepackage{amsfonts}
\usepackage{amsmath}
\usepackage{eurosym}

\begin{document}

\title{Entropic Dynamics: Reconstructing Quantum Field Theory in Curved
Space-time}
\author{Selman Ipek, Mohammad Abedi, and Ariel Caticha \\
{\small Physics Department, University at Albany-SUNY, Albany, NY 12222, USA.%
}}
\date{}
\maketitle

\begin{abstract}
The Entropic Dynamics reconstruction of quantum mechanics is extended to the
quantum theory of scalar fields in curved space-time. The Entropic Dynamics
framework, which derives quantum theory as an application of the method of
maximum entropy, is combined with the covariant methods of Dirac, Hojman,
Kucha\v{r}, and Teitelboim, which they used to develop a framework for
classical covariant Hamiltonian theories. The goal is to formulate an
information-based alternative to current approaches based on algebraic
quantum field theory. One key ingredient is the adoption of a local notion
of entropic time in which instants are defined on curved three-dimensional
surfaces and time evolution consists of the accumulation of changes induced
by local deformations of these surfaces. The resulting dynamics is a
non-dissipative diffusion that is constrained by the requirements of
foliation invariance and incorporates the necessary local quantum
potentials. As applications of the formalism we derive the Ehrenfest relations for
fields in curved-spacetime and briefly discuss the nature of divergences in
quantum field theory. 
\end{abstract}

\section{Introduction}

Research on quantum field theory in curved space-time (QFTCS) has been the
subject of a sustained effort for several decades. (See, for example, \cite%
{Birrell Davies 1984}\cite{Wald 1994} and the recent review \cite{Hollands
Wald 2014}.) The motivation is twofold. There is an intrinsic interest in
QFTCS as an effective theory that is reliable except in those regions of
space-time where the curvature is extremely high. But there is also a
widespread belief that QFTCS is a necessary intermediate step on the way
towards a more fundamental theory of interacting quantum and gravitational
fields.

As discussed in \cite{Hollands Wald 2014} QFTCS is usually formulated by
merging classical general relativity with quantum field theory. Identifying
the principles that govern the gravity sector does not appear to be
problematic. Unfortunately, the choice of which principles of quantum field
theory in flat space-time should be retained in a curved space-time is not
nearly as clear. The lack of Poincare invariance implies the absence of a
preferred vacuum state and forces one to revisit the notion of particle \cite%
{Fulling 1973}\cite{Unruh 1976}. The representation of quantum fields as
operators, or better, as operator-valued distributions, is problematic on
many levels. Prominent among them is the issue of the Hilbert space itself.
A representation of the relevant commutation relations and field equations
requires operators that act on a Hilbert space. The problem is that in
curved space-time there is an infinite number of such representations that
are unitarily inequivalent. Another problem is that representing fields as
operators may itself be deeply misguided --- it may turn out to be an
attempt to preserve in curved space something that has already turned out to
be questionable at the level of flat space-time, namely, the standard
quantum theory of measurement.

But there is another direction from which this subject can be approached.
The discoveries of black hole entropy and thermodynamics \cite{Bekenstein
1973}-\cite{Hawking 1976} suggest a deep connection between the fundamental
laws of physics and information. Indeed, in recent decades the subject of
foundations of quantum mechanics (FQM) has undergone a renaissance (for
reviews see \emph{e.g. }\cite{Schlosshauer 2004}-\cite{Leifer 2014})
culminating in a variety of proposals for the reconstruction of quantum
mechanics (see \emph{e.g.}, \cite{Nelson 1985}-\cite{tHooft 2016}). Notable
among the latter are proposals that seek a deeper foundation based on
information theory (see \emph{e.g.}, \cite{Wootters 1981}-\cite{Reginatto
Hall 2012}). To a large extent the two subjects --- QFTCS and FQM --- have
developed independently. In fact, it is somewhat surprising that progress in
QFTCS has contributed so little towards clarifying the conceptual
foundations of quantum mechanics and vice versa.

The purpose of this paper is to contribute to bridge this gap by extending
one particular approach to the reconstruction of quantum mechanics --- the
entropic dynamics approach --- to the reconstruction of quantum field theory.

Entropic Dynamics (ED) is a framework for constructing dynamical theories on
the basis of Bayesian and entropic principles of inference \cite{Caticha
2010}\cite{Caticha 2015}. (For a recent self-contained presentation and
review see \cite{Caticha 2017}.) The goal is to develop models for the time
evolution of the probability distributions of the positions of particles or
the values of fields. ED differs from other information approaches to
quantum theory by strictly adhering to Bayesian and entropic methods of
inference, and by paying close attention to the notion of time.

One challenge that ED was designed to address is that a fully epistemic%
\footnote{%
Within a model such as ED a concept is referred as `epistemic' when it is
related to the state of knowledge, opinion, or belief of a rational agent.
For example, Bayesian probabilities are epistemic --- they are tools for
reasoning with incomplete information \cite{Caticha 2012}. } interpretation
of the wave function $\Psi $ is not achieved by merely declaring that the
square of a wave function $|\Psi |^{2}$ yields a Bayesian probability. To
insure consistency one must also show that the rules for updating those
probabilities --- which include both the unitary evolution of the wave
function $\Psi $ and its \textquotedblleft collapse\textquotedblright\
during a measurement --- are in strict accord with Bayesian and entropic
methods \cite{Jaynes 1957}-\cite{Caticha 2012}. Thus, in an \emph{entropic}
dynamics the evolution must be driven by information codified into
constraints; it is through these constraints that the \textquotedblleft
physics\textquotedblright\ is introduced.

An important early insight in the context of Nelson's stochastic mechanics,
was his realization that important aspects of quantum mechanics could be
modelled as a non-dissipative Brownian motion \cite{Nelson 1979}. This idea
was suitably adapted to the ED setting and imposing the conservation of an
appropriate energy functional became the main criterion for choosing the
evolving constraints \cite{Caticha 2010}. It eventually led to a fully
Hamiltonian formalism with its attendant action principle, a symplectic
structure, and Poisson brackets \cite{Bartolomeo et al 2014}\cite{Caticha
2017}. Unfortunately in curved space-times the energy criterion is not
satisfactory because the notion of a \emph{global} energy is not in general
available. An alternative \emph{local} criterion for evolving constraints is
needed.

Another challenge that ED is designed to address concerns the concept of
time. Entropic time is constructed as a scheme to keep track of the
accumulation of small changes \cite{Caticha 2010}. This involves identifying
a suitable notion of ordered \textquotedblleft instants\textquotedblright\
and then introducing a convenient measure of the interval or separation
between them. In ED an instant is defined through the information that is
necessary to predict or \textquotedblleft construct\textquotedblright\ the
next instant. This amounts to specifying a state of knowledge together with
a purely kinematic criterion of simultaneity. In \cite{Caticha 2012b}\cite%
{Ipek Caticha 2014} the concept of a \emph{global} instant was used to
generate an ED of quantum scalar fields in Minkowski space-time. However,
although the model was fully relativistic, its relativistic invariance was
not manifest, that is, the freedom to represent the relativity of
simultaneity was not explicit.

The goal of this paper is to formulate an inference-based approach to QFTCS
in which quantum fields are not represented as operators; there is no
mention of a preferred vacuum state, nor of an associated Hilbert space.%
\footnote{%
A brief presentation of some of these ideas appeared in \cite{Ipek et al
2017}.} We construct a manifestly relativistic quantum ED in curved
space-time incorporating ideas developed in the classical field theories of
Dirac, Hojman, Kucha\v{r}, and Teitelboim (DHKT) \cite{Dirac 1951}-\cite%
{Hojman et al 1976} which can themselves be traced to the earlier theories
of Weiss, Tomonaga, Dirac, and Schwinger \cite{Weiss 1938}-\cite{Schwinger
1948}. Drawing on the ideas of DHKT, we relax the assumption of a global
time in favor of a notion of local time, and the non-covariant criterion of
energy conservation is replaced by the covariant requirement of foliation
invariance. In this view of ED, an instant is defined by a three-dimensional
spacelike surface embedded in space-time plus the fields and probability
distributions defined on such surfaces. In a fully covariant theory, such
surfaces are constructed by slicing or foliating space-time into a sequence
of spacelike surfaces. The freedom to choose the foliation, which amounts to
the local relativity of simultaneity, is implemented by a consistency
requirement: the evolution of all dynamical quantities from an initial to a
final surface must be independent of the choice of intermediate surfaces. We
can refer to foliation invariance either as a requirement of consistency or,
following Kucha\v{r} and Teitelboim, as a requirement of \textquotedblleft
path independence\textquotedblright\ --- if there are two alternative ways
to evolve from an initial to a final instant, then the two ways must lead to
the same result.\footnote{%
Wheeler originally called this an \textquotedblleft
embeddability\textquotedblright\ condition --- the requirement that all
these surfaces be embedded in the same space-time.}

The formalism is developed for a scalar field $\chi (x)$. First we find the
transition probability for an infinitesimal change --- the effect of an
infinitesimal local deformation of the instant. The introduction of a local
entropic time then leads to a set of functional \textquotedblleft local-time
Fokker-Planck\textquotedblright\ diffusion equations for the evolution of
probability distribution $\rho \lbrack \chi ]$. Requiring that the evolution
satisfy foliation invariance leads to a non-dissipative Hamiltonian
diffusion. This is done by imposing that the generators that update the
entropic constraints satisfy the same DHKT \textquotedblleft
algebra\textquotedblright\ as the generators that describe the kinematics of
surface deformations. The result is a covariant quantum entropic dynamics of
scalar fields in curved space-time.

This paper focuses on quantun field theory but the ED approach has been
applied to a variety of other topics in quantum theory. These include: the
quantum measurement problem \cite{Johnson Caticha 2011}\cite{Vanslette
Caticha 2016}; momentum and uncertainty relations \cite{Nawaz Caticha 2011};
extensions to curved spaces \cite{Nawaz et al 2015}; the Bohmian limit \cite%
{Bartolomeo Caticha 2015}\cite{Bartolomeo Caticha 2016}; the classical limit 
\cite{Demme Caticha 2016};  and the ED of spin \cite{Caticha Carrara 2019}.

\section{The Entropic Dynamics of infinitesimal steps}

\paragraph{The microstates---}

In an inference scheme such as ED the physics is introduced through the
choice of dynamical variables and of the constraints that capture the
dynamically relevant information. Here we consider the dynamics of a scalar
field $\chi \left( x\right) $. For notational convenience we will often
write the $x$-dependence as a subscript, $\chi \left( x\right) =\chi _{x}$.
Unlike the standard Copenhagen interpretation of quantum theory where
observables have definite values only when elicited through an experiment,
in the ED approach these fields have definite values at all times. However,
these values are unknown and the dynamics is indeterministic.

The field $\chi $ lives on a $3$-dimensional curved space $\sigma $, the
points of which are labeled by coordinates $x^{i}$ ($i=1,2,3$). The space $%
\sigma $ is endowed with a metric $g_{ij}$ induced by the non-dynamical
background space-time in which $\sigma $ is embedded. Thus $\sigma $ is an
embedded hypersurface; for simplicity we shall refer to it as a
\textquotedblleft surface\textquotedblright . We assume that space-time is
globally hyperbolic. It can be foliated by Cauchy surfaces and its topology
is $\mathbb{R}\times \Sigma $ where $\Sigma $ stands for any Cauchy surface $%
\sigma $ \cite{Wald 1984}. The field $\chi _{x}$ is a scalar with respect to
3-diffeomorphisms on the surface $\sigma $. The $\infty $-dimensional space
of all possible field configurations is the configuration space $\mathcal{C}$%
. A single field configuration, labelled $\chi $, is represented as a point
in $\mathcal{C}$, and the uncertainty in the field is described by a
probability distribution $\rho \lbrack \chi ]$ over $\mathcal{C}$.

\paragraph{Maximum Entropy---}

Our goal here is to predict the evolution of the scalar field $\chi $. To
this end we make one major assumption: in ED, motion is continuous, the
fields follow continuous trajectories in $\mathcal{C}$. This implies that
any finite change can be analyzed as the accumulation of many
infinitesimally short steps. Therefore our first goal is to calculate the
probability $P\left[ \chi ^{\prime }|\chi \right] $ that the field undergoes
a small change from an initial configuration $\chi $ to a neighboring $\chi
^{\prime }=\chi +\Delta \chi $ and later we calculate the probability of the
finite change that results from a sequence of short steps. The transition
probability $P\left[ \chi ^{\prime }|\chi \right] $ is found by maximizing
the entropy functional, 
\begin{equation}
S\left[ P,Q\right] =-\int D\chi ^{\prime }P\left[ \chi ^{\prime }|\chi %
\right] \log \frac{P\left[ \chi ^{\prime }|\chi \right] }{Q\left[ \chi
^{\prime }|\chi \right] },  \label{entropy a}
\end{equation}%
relative to a prior $Q\left[ \chi ^{\prime }|\chi \right] $ and subject to
appropriate constraints. It is through the prior and the constraints that
the relevant physical information is introduced.

\paragraph{The prior---}

We adopt a prior $Q\left[ \chi ^{\prime }|\chi \right] $ that incorporates
the information that change happens by infinitesimally small amounts, but is
otherwise maximally uninformative. In particular, as far as the prior is
concerned, changes at different points are uncorrelated. Such a prior can
itself be derived from the principle of maximum entropy. Indeed, maximize 
\begin{equation}
S[Q,\mu ]=-\int d\chi ^{\prime }\,Q\left[ \chi ^{\prime }|\chi \right] \log 
\frac{Q\left[ \chi ^{\prime }|\chi \right] }{\mu (\chi ^{\prime })}~,
\label{entropy b}
\end{equation}%
relative to the measure $\mu (\chi ^{\prime })$ which we assume to be
uniform and subject to appropriate constraints.\footnote{%
Since we deal with infinitesimally short steps the prior $Q$ turns to be
quite independent of the choice of the underlying measure $\mu $.} The
requirement that the field undergoes changes that are small and uncorrelated
is implemented by imposing an infinite number of independent constraints,
one at each point $x$, 
\begin{equation}
\langle \Delta \chi _{x}^{2}\rangle =\int D\chi ^{\prime }\,Q\left[ \chi
^{\prime }|\chi \right] (\Delta \chi _{x})^{2}=\kappa _{x}\,,
\label{Constraint 1}
\end{equation}%
where $\Delta \chi _{x}=\chi _{x}^{\prime }-\chi _{x}$ and, to enforce the
continuity of the motion, we will eventually take the limit $\kappa
_{x}\rightarrow 0$. The result of maximizing (\ref{entropy b}) subject to (%
\ref{Constraint 1}) and normalization is a product of Gaussians, 
\begin{equation}
Q\left[ \chi ^{\prime }|\chi \right] \propto \,\exp -\frac{1}{2}\int
dx\,g_{x}^{1/2}\alpha _{x}\left( \Delta \chi _{x}\right) ^{2}~  \label{prior}
\end{equation}%
where $\alpha _{x}$ are the Lagrange multipliers associated to each
constraint (\ref{Constraint 1}), and the scalar density $g_{x}^{1/2}=\,%
\left( \det \,g_{ij}\right) ^{1/2}$ is introduced so that $\alpha _{x}$ is a
scalar field. The limit $\kappa _{x}\rightarrow 0$ is achieved by taking $%
\alpha _{x}\rightarrow \infty $. For notational simplicity we write $dx$
instead of $d^{3}x$.

\paragraph*{The drift potential constraint---}

The dynamics induced by the prior (\ref{prior}) is a diffusion that is
isotropic in configuration space. In order to introduce correlations,
directionality, and such quintessential quantum effects as interference and
entanglement, we impose one additional \emph{single} constraint that is 
\emph{non-local} in space but local in configuration space. This single
constraint involves a functional on configuration space, the
\textquotedblleft drift\textquotedblright\ potential $\phi \lbrack \chi ]$.
We impose that the expectation of the change of the drift potential $\Delta
\phi =\phi \left[ \chi ^{\prime }\right] -\phi \left[ \chi \right] $ is
another small quantity $\kappa ^{\prime }$ that will eventually be taken to
zero, 
\begin{equation}
\langle \Delta \phi \rangle =\kappa ^{\prime }\quad \text{or}\quad \int_{%
\mathcal{C}}D\chi ^{\prime }\,P\left[ \chi ^{\prime }|\chi \right] \int
dx\,\,\Delta \chi _{x}\frac{\delta \phi \left[ \chi \right] }{\delta \chi
_{x}}=\kappa ^{\prime }.  \label{Constraint 2}
\end{equation}%
(Note that since $\chi _{x}$ and $\Delta \chi _{x}$ are scalars, in order
for (\ref{Constraint 2}) to be invariant under coordinate transformations of
the surface the derivative $\delta /\delta \chi _{x}$ must transform as a
scalar density.) The physical meaning of the drift potential $\phi \lbrack
\chi ]$ will not be discussed here. As in so many other situations in
physics the mere identification of forces and constraints can turn out to be
useful even when their microscopic origins is not yet fully understood.%
\footnote{%
Elsewhere we show that in order to describe single particle states with
nonzero angular momentum $\phi $ needs to have the topological properties of
an angle \cite{Ipek 2019}. Additional evidence that $\phi $ must be
interpreted as an angle is provided in the ED of non-relativistic particles
with spin \cite{Caticha Carrara 2019}.}

\paragraph*{The transition probability---}

Next we maximize (\ref{entropy a}) subject to (\ref{Constraint 2}) and
normalization. As discussed in \cite{Bartolomeo Caticha 2016} the multiplier 
$\alpha ^{\prime }$ associated to the global constraint (\ref{Constraint 2})
turns out to have no influence on the dynamics: it can be absorbed into the
drift potential $\alpha ^{\prime }\phi \rightarrow \phi $ \ which means we
can effectively set $\alpha ^{\prime }=1$. The resulting transition
probability is a Gaussian distribution, 
\begin{equation}
P\left[ \chi ^{\prime }|\chi \right] =\frac{1}{Z\left[ \alpha _{x},g_{x}%
\right] }\,\exp -\frac{1}{2}\int dx\,g_{x}^{1/2}\alpha _{x}\left( \Delta
\chi _{x}-\frac{1}{g_{x}^{1/2}\alpha _{x}}\frac{\delta \phi \left[ \chi %
\right] }{\delta \chi _{x}}\right) ^{2},  \label{Trans Prob}
\end{equation}%
where $Z\left[ \alpha _{x},g_{x}\right] $ is the normalization constant. In
previous work \cite{Caticha 2012b}\cite{Ipek Caticha 2014} $\alpha _{x}$ was
chosen to be a spatial constant $\alpha $ to reflect the translational
symmetry of flat space. Here we make no such restriction and instead relax
the global constant $\alpha $ in favor of a non-uniform spatial scalar $%
\alpha _{x}$ which will be a key element in implementing our scheme for a
local entropic time.

The Gaussian form of (\ref{Trans Prob}) allows us to present a generic
change, 
\begin{equation}
\Delta \chi _{x}=\left\langle \Delta \chi _{x}\right\rangle +\Delta w_{x}~,
\end{equation}
as resulting from an expected drift $\left\langle \Delta \chi
_{x}\right\rangle $ plus fluctuations $\Delta w_{x}$. At each $x$ the
expected short step is 
\begin{equation}
\left\langle \Delta \chi _{x}\right\rangle =\frac{1}{g_{x}^{1/2}\,\alpha _{x}%
}\frac{\delta \phi \left[ \chi \right] }{\delta \chi _{x}}\equiv \Delta \bar{%
\chi}_{x},  \label{Exp Step 1}
\end{equation}%
while the fluctuations $\Delta w_{x}$ satisfy,%
\begin{equation}
\left\langle \Delta w_{x}\right\rangle =0\,,\quad \text{and}\hspace{0.4cm}%
\left\langle \Delta w_{x}\Delta w_{x^{\prime }}\right\rangle =\frac{1}{%
g_{x}^{1/2}\alpha _{x}}\delta _{xx^{\prime }}.  \label{Fluctuations}
\end{equation}%
Thus we see that $\Delta \bar{\chi}_{x}\sim 1/\alpha _{x}$ and $\Delta
w_{x}\sim 1/\alpha _{x}^{1/2}$, so that for short steps, $\alpha
_{x}\rightarrow \infty $, the fluctuations dominate the motion. The
resulting trajectory is continuous but non-differentiable --- a Brownian
motion.

\section{Entropic time}

In ED the idea of time is derived from the idea of change. Time is
introduced as a device to keep track of how the accumulation of many
infinitesimal changes builds up into a finite change. Questions such as,
\textquotedblleft What is an instant?\textquotedblright\ \textquotedblleft
How are they ordered?\textquotedblright\ and \textquotedblleft To what
extent are they separated?\textquotedblright\ are central to constructing
any dynamical theory and ED is no exception.

\paragraph{Ordered instants---}

Of particular importance is the notion of an instant, which in ED involves
several ingredients: (1) A foliation of spacelike surfaces $\sigma $ that
codify spatial relations and provide a criterion of simultaneity and
duration. (2) We must specify the \textquotedblleft epistemic
contents\textquotedblright\ of the surfaces. This is a specification of a
statistical state that is sufficient for the prediction of future states. It
is given by a probability distribution $\rho \lbrack \chi ]$ and a drift
potential $\phi \lbrack \chi ]$. And (3) an entropic step in which the
statistical state at one instant is updated to generate the state at the
next instant. This is the requirement that generates the sequence of ordered
instants which makes the dynamics come alive.

\paragraph*{Some space-time kinematics---}

We deal with a curved space-time; events are labeled by space-time
coordinates $X^{\mu }$; and the metric is $g_{\mu \nu }\left( X^{\beta
}\right) $.\footnote{%
We use Greek indices ($\mu ,\nu ,...$ $=0,1,2,3$) for space-time coordinates 
$X^{\mu }$ and latin indices ($a,b,...i,j,...$ $=1,2,3$) for coordinates $%
x^{i}$ on the surface $\sigma $. The spacetime metric has signature $(-+++)$.%
} Space-time is foliated by a sequence of space-like surfaces $\left\{
\sigma \right\} $. Points on the surface $\sigma $ are labeled by
coordinates $x^{i}$ and the embedding of the surface within space-time is
defined by four functions $X^{\mu }=X^{\mu }\left( x^{i}\right) $. The
metric induced on the surface is%
\begin{equation}
g_{ij}\left( x\right) =X_{i}^{\mu }X_{j}^{\nu }g_{\mu \nu }\quad \text{where}%
\quad X_{i}^{\mu }=\frac{\partial X^{\mu }}{\partial x^{i}}\ .
\label{induced metric}
\end{equation}%
The metric $g_{ij}$ will in general depend on the particular surface. In
this work neither $g_{\mu \nu }\left( X\right) $ nor $g_{ij}\left( x\right) $
are themselves dynamical.

Following Teitelboim and Kucha\u{r}, we consider an infinitesimal
deformation of the surface $\sigma $ to a neighboring surface $\sigma
^{\prime }$. This is specified by the deformation vector $\delta \xi ^{\mu }$
which connects the point in $\sigma $ with coordinates $x^{i}$ to the point
in $\sigma ^{\prime }$ with the same coordinates $x^{i}$, 
\begin{equation}
\delta \xi ^{\mu }=\delta \xi ^{\bot }n^{\mu }+\delta \xi ^{i}X_{i}^{\mu }~,
\label{deformation vector}
\end{equation}%
where $n^{\mu }$ is the unit normal to the surface ($n_{\mu }n^{\mu }=-1$, $%
n_{\mu }X_{i}^{\mu }=0$). The normal and tangential components are given by%
\begin{equation}
\delta \xi _{x}^{\bot }=-n_{\mu x}\delta \xi _{x}^{\mu }\quad \text{and}%
\quad \delta \xi _{x}^{i}=X_{\mu x}^{i}\delta \xi _{x}^{\mu }~,
\end{equation}%
where $X_{\mu x}^{i}=g^{ij}g_{\mu \nu }X_{jx}^{\nu }$. They are known as
lapse and shift respectively and are collectively denoted ${}(\delta \xi
^{\bot },\delta \xi ^{i})=\delta \xi ^{A}$. As a matter of convention a
deformation is identified by its normal $\delta \xi ^{\bot }$ and tangential 
$\delta \xi ^{i}$ components independently of the surface upon which it acts
(\emph{i.e.}, independently of the normal $n^{\mu }$). This allows us to
speak about applying \emph{the same deformation} to different surfaces; a
useful concept for our discussion of path independence.

\paragraph{Duration---}

In ED time is defined so that motion looks simple. The specification of the
time interval between two successive instants is dictated purely by
convenience, that is, the best choice is that which reflects the various
symmetries of the physical situation. For example, in a non-relativistic QM
one would adopt a Newtonian time $\Delta t$ that reflects the translational
symmetries of space so that \textquotedblleft time flows equably everywhere
and everywhen\textquotedblright . In our case the most convenient choice is
one that reflects the symmetries of the background curved space-time.

Since for short steps the dynamics is dominated by fluctuations, eq.(\ref%
{Fluctuations}), the choice of time interval is achieved through an
appropriate choice of the multipliers $\alpha _{x}$. So far the present
development of ED has followed closely along the lines of the non-covariant
models discussed in \cite{Bartolomeo et al 2014} and \cite{Ipek Caticha 2014}%
). The important point of departure is that here we are concerned with
instants defined on the curved embedded surfaces $\sigma $ and $\sigma
^{\prime }$. It is then natural to define a local notion of duration in
terms of an invariant --- the proper time. The idea is the familiar one: at
the point $x$ in $\sigma $ draw a normal segment reaching out to $\sigma
^{\prime }$. \emph{The proper time }$\delta \xi _{x}^{\bot }$\emph{\ along
this normal segment provides us with the local measure of duration between }$%
\sigma $\emph{\ and }$\sigma ^{\prime }$\emph{\ at }$x$\emph{.} More
specifically, let 
\begin{equation}
\alpha _{x}=\frac{1}{\delta \xi _{x}^{\bot }}\quad \text{so that}\quad
\left\langle \Delta w_{x}\Delta w_{x^{\prime }}\right\rangle =\frac{\,\delta
\xi _{x}^{\bot }}{g_{x}^{1/2}}\delta _{xx^{\prime }}~.  \label{Duration}
\end{equation}

\paragraph{The statistical state and its evolution---}

Entropic dynamics is generated by the short-step transition probability $P%
\left[ \chi ^{\prime }|\chi \right] $. In a generic short step both the
initial $\chi $ and the final $\chi ^{\prime }$ are unknown. Integrating the
joint probability, $P\left[ \chi ^{\prime },\chi \right] $, over $\chi $
gives 
\begin{equation}
P[\chi ^{\prime }]=\int d\chi \,P\left[ \chi ^{\prime },\chi \right] =\int
d\chi P\left[ \chi ^{\prime }|\chi \right] P\left[ \chi \right] ~.
\label{CK a}
\end{equation}%
These equations are true by virtue of the laws of probability; they involve
no assumptions. However, if $P\left[ \chi \right] $ happens to be the
probability of $\chi $ \emph{at an \textquotedblleft
instant\textquotedblright\ labelled }$\sigma $, then we can interpret $P%
\left[ \chi ^{\prime }\right] $ as the probability of $\chi ^{\prime }$ 
\emph{at the \textquotedblleft next instant,\textquotedblright } which we
will label $\sigma ^{\prime }$. Accordingly, we write $P\left[ \chi \right]
=\rho _{\sigma }\left[ \chi \right] $ and $P\left[ \chi ^{\prime }\right]
=\rho _{\sigma ^{\prime }}\left[ \chi ^{\prime }\right] $ so that%
\begin{equation}
\rho _{\sigma ^{\prime }}\left[ \chi ^{\prime }\right] =\int D\chi \,P\left[
\chi ^{\prime }|\chi \right] \,\rho _{\sigma }\left[ \chi \right] ~.
\label{Evolution equation}
\end{equation}%
This is the basic dynamical equation; it allows one to update the
statistical state $\rho _{\sigma }\left[ \chi \right] $ from one instant to
the next. Note that since $P\left[ \chi ^{\prime }|\chi \right] $ is found
by maximizing entropy not only are these instants ordered but there is a
natural \emph{entropic} arrow of time: $\sigma ^{\prime }$ occurs \emph{after%
} $\sigma $. But we are not done yet. With the definition (\ref{Duration})
of duration, the dynamics given by (\ref{Evolution equation}) and (\ref%
{Trans Prob}) describes a Wiener process evolving along a given foliation of
space-time. To obtain a fully covariant dynamics we require that the
evolution of any dynamical quantity such as $\rho _{\sigma }[\chi ]$ from an
initial $\sigma _{i}$ to a final $\sigma _{f}$ must be independent of the
intermediate choice of surfaces. This \textquotedblleft foliation
invariance\textquotedblright\ or \textquotedblleft path
independence\textquotedblright , which amounts to the local relativity of
simultaneity, is a consistency requirement: if there are different ways to
evolve from a given initial instant into a given final instant, then all
these ways must agree. The conditions to implement this consistency are the
subject of the next section.

\paragraph{The local-time diffusion equations---}

The dynamics expressed in integral form by (\ref{Evolution equation}) with (%
\ref{Trans Prob}) and (\ref{Duration}) can be rewritten in differential form
as an infinite set of local equations, one for each spatial point, 
\begin{equation}
\frac{\delta \rho _{\sigma }}{\delta \xi _{x}^{\bot }}=-\,g_{x}^{-1/2}\frac{%
\delta }{\delta \chi _{x}}\left( \rho _{\sigma }\,\frac{\delta \Phi _{\sigma
}}{\delta \chi _{x}}\right) \hspace{0.5cm}\text{with}\hspace{0.5cm}\Phi
_{\sigma }\left[ \chi \right] =\,\phi _{\sigma }\left[ \chi \right] -\log
\rho _{\sigma }^{1/2}\left[ \chi \right] ~.  \label{FP equation}
\end{equation}%
(The derivation is given in Appendix \ref{appendix FP}.) This set of
equations describes the flow of the probability $\rho _{\sigma }\left[ \chi %
\right] $ in the configuration space $\mathcal{C}$ as the surface $\sigma $
is deformed. More explicitly, the actual change in $\rho \lbrack \chi ]$ as $%
\sigma $ is infinitesimally deformed to $\sigma ^{\prime }$ is 
\begin{equation}
\delta _{\bot }\rho _{\sigma }\left[ \chi \right] =\int dx\frac{\delta \rho
_{\sigma }\left[ \chi \right] }{\delta \xi _{x}^{\bot }}\delta \xi
_{x}^{\bot }=-\int dx\frac{\delta \xi _{x}^{\bot }}{g_{x}^{1/2}}\frac{\delta 
}{\delta \chi _{x}}\left( \rho _{\sigma }\left[ \chi \right] \frac{\delta
\Phi _{\sigma }\left[ \chi \right] }{\delta \chi _{x}}\right) \,.
\label{FP b}
\end{equation}%
In the special case when both surfaces $\sigma $ and $\sigma ^{\prime }$
happen to be flat then $g_{x}^{1/2}=1$ and $\delta \xi _{x}^{\bot }=dt$ are
constants and eq.(\ref{FP b}) becomes 
\begin{equation}
\frac{\partial \rho _{t}\left[ \chi \right] }{\partial t}=-\int dx\frac{%
\delta }{\delta \chi _{x}}\left( \rho _{t}\left[ \chi \right] \frac{\delta
\Phi _{t}\left[ \chi \right] }{\delta \chi _{x}}\right) \,,  \label{FP c}
\end{equation}%
which we recognize as a diffusion or Fokker-Planck equation written as a
continuity equation for the flow of probability in configuration space $%
\mathcal{C}$. Accordingly we will refer to (\ref{FP equation}) as the
\textquotedblleft local-time Fokker-Planck\textquotedblright\ equations
(LTFP). These equations describe the flow of probability with a current
velocity $v_{x}\left[ \chi \right] =\delta \Phi /\delta \chi _{x}$\thinspace
. Eventually, the functional $\Phi $ will be identified as the
Hamilton-Jacobi functional, or the phase of the wave functional in the
quantum theory.

Anticipating later developments we note that the LTFP eqs.(\ref{FP equation}%
) can be rewritten in an alternative and very suggestive form involving the
notion of an ensemble functional or \emph{e-functional}. Just as a regular
functional such as $\rho \left[ \chi \right] $ maps a field distribution $%
\chi $ into a real number (a probability in this case), an e-functional maps
a functional, such as $\rho \left[ \chi \right] $ or $\Phi \left[ \chi %
\right] $, into a real number. Then, just as one can define functional
derivatives, one can also define e-functional derivatives.\footnote{%
An excellent brief review of the ensemble calculus is given in the appendix
of \cite{Hall et al 2003}.} Introduce an e-functional $\tilde{H}_{\perp x}%
\left[ \rho _{\sigma },\Phi _{\sigma }\right] $ such that at each point $x$ 
\begin{equation}
\frac{\delta \rho _{\sigma }\left[ \chi \right] }{\delta \xi _{x}^{\bot }}=%
\frac{\tilde{\delta}\tilde{H}_{\perp x}\left[ \rho _{\sigma },\Phi _{\sigma }%
\right] }{\tilde{\delta}\Phi _{\sigma }\left[ \chi \right] }
\label{FP equation H}
\end{equation}%
reproduces (\ref{FP equation}). In what follows we denote all ensemble
quantities such as $\tilde{H}_{\perp x}$ with a tilde: $\tilde{\delta}/%
\tilde{\delta}\Phi \lbrack \chi ]$ is the e-functional derivative with
respect to $\Phi \left[ \chi \right] $. We stress that writing (\ref{FP
equation}) in the form (\ref{FP equation H}) does not involve any new
assumptions; an appropriate $\tilde{H}_{\perp x}$ can always be found.
Indeed, substitute (\ref{FP equation}) into the left of (\ref{FP equation H}%
), then an easy integration gives

\begin{equation}
\tilde{H}_{\perp x}\left[ \rho ,\Phi ;\sigma ,\chi \right] =\int D\chi
\,\rho \frac{1}{2g_{x}^{1/2}}\left( \frac{\delta \Phi }{\delta \chi _{x}}%
\right) ^{2}+F_{x}\left[ \rho ;\sigma ,\chi \right] \,,  \label{e-Hp a}
\end{equation}%
where the integration constants $F_{x}=F_{x}[\rho ;\sigma ,\chi ]$ are
independent of $\Phi $; they may depend on $\rho $, on the geometry of the
surface $\sigma $, and also on the fields $\chi _{x}$.\footnote{%
The notation $F_{x}[\rho ;\sigma ,\chi ]$ indicates that $F_{x}$ is a
functional of various dynamical and non-dynamical variables. To unclutter
the notation some of these dependencies will not be explicitly displayed.}
In later sections we will see that $\tilde{H}_{\perp x}$ captures dynamical
information about the evolution of $\Phi $ as well as $\rho $ and can be
cast as a \textit{Hamiltonian} generator.\footnote{%
Eq.(\ref{FP equation H}) provides the criterion that allows us to identify
the momentum $\Phi \lbrack \chi ]$ that is canonically conjugated to the
generalized coordinate $\rho \lbrack \chi ]$.}

\section{Kinematics of surface deformations and their generators}

Dynamics in local time must reflect the kinematics of surface deformations,
and this kinematics can be studied independently of the particular dynamics
being considered. As a surface is deformed, its geometry and, more
generally, the statistical state associated with it is also subject to
change. Consider a generic functional $T\left[ X(x)\right] $ that assigns a
number to every surface $X^{\mu }(x)$. The change in the functional $\delta
T $ resulting from an arbitrary deformation $\delta \xi _{x}^{A}$ has the
form 
\begin{equation}
\delta T=\int dx\,\delta \xi _{x}^{\mu }\frac{\delta T}{\delta \xi _{x}^{\mu
}}=\int dx\,\left( \delta \xi _{x}^{\bot }G_{\bot x}+\delta \xi
_{x}^{i}G_{ix}\right) T~\,,  \label{delta T}
\end{equation}%
where 
\begin{equation}
G_{\bot x}=\frac{\delta }{\delta \xi _{x}^{\bot }}=n_{x}^{\mu }\frac{\delta 
}{\delta X_{x}^{\mu }}\quad \text{and}\quad G_{ix}=\frac{\delta }{\delta \xi
_{x}^{i}}=X_{ix}^{\mu }\frac{\delta }{\delta X_{x}^{\mu }}
\end{equation}%
are the generators of normal and tangential deformations respectively.
Unlike the vectors, $\delta /\delta X_{x}^{\mu }$, which form a coordinate
basis in the space of hypersurfaces and therefore commute, the generators of
deformations $\delta /\delta \xi _{x}^{A}$ form a non-holonomic basis. Their
non-vanishing commutator is 
\begin{equation}
\frac{\delta }{\delta \xi _{x}^{A}}\frac{\delta }{\delta \xi _{x^{\prime
}}^{B}}-\frac{\delta }{\delta \xi _{x^{\prime }}^{B}}\frac{\delta }{\delta
\xi _{x}^{A}}=\int dx^{\prime \prime }\,\kappa ^{C}{}_{BA}(x^{\prime \prime
};x^{\prime },x)\frac{\delta }{\delta \xi _{x^{\prime \prime }}^{C}}
\label{commutator}
\end{equation}%
where $\kappa ^{C}{}_{BA}$ are the \textquotedblleft structure
constants\textquotedblright\ of the \textquotedblleft
group\textquotedblright\ of deformations.

The calculation of $\kappa ^{C}{}_{BA}$ is given in \cite{Teitelboim 1973a}%
\cite{Kuchar 1973}. The basic idea is \emph{embeddability}: When we perform
two successive infinitesimal deformations $\delta \xi ^{A}$ followed by $%
\delta \eta ^{B}$, $\sigma \overset{\delta \xi }{\rightarrow }\sigma _{1}%
\overset{\delta \eta }{\rightarrow }\sigma ^{\prime }$, the three surfaces
are all embedded in the same space-time. The same happens when we execute
them in the opposite order, $\sigma \overset{\delta \eta }{\rightarrow }%
\sigma _{2}\overset{\delta \xi }{\rightarrow }\sigma ^{\prime \prime }$. The
key point is that the since the surfaces $\sigma ^{\prime }$ and $\sigma
^{\prime \prime }$ are embedded in the same space-time there must exist a
third deformation $\delta \zeta ^{\alpha }$ that takes $\sigma ^{\prime }$
to $\sigma ^{\prime \prime }$: $\sigma ^{\prime }\overset{\delta \zeta }{%
\rightarrow }\sigma ^{\prime \prime }$. Thus the deformation from $\sigma $
to $\sigma ^{\prime \prime }$ can be attained by following two different
paths: either we follow the direct path $\sigma \overset{\delta \eta }{%
\rightarrow }\sigma _{2}\overset{\delta \xi }{\rightarrow }\sigma ^{\prime
\prime }$ or we follow the indirect path $\sigma \overset{\delta \xi }{%
\rightarrow }\sigma _{1}\overset{\delta \eta }{\rightarrow }\sigma ^{\prime }%
\overset{\delta \zeta }{\rightarrow }\sigma ^{\prime \prime }$. Then, as
shown in \cite{Teitelboim 1973a}\cite{Kuchar 1973}, eq.(\ref{commutator})
leads to the \textquotedblleft algebra\textquotedblright , 
\begin{eqnarray}
\lbrack G_{\bot x},G_{\bot x^{\prime }}] &=&-(g_{x}^{ij}G_{jx}+g_{x^{\prime
}}^{ij}G_{jx^{\prime }})\partial _{ix}\delta (x,x^{\prime })~,  \label{LB1}
\\
\lbrack G_{ix},G_{\bot x^{\prime }}] &=&-G_{\bot x}\partial _{ix}\delta
(x,x^{\prime })~,  \label{LB2} \\
\lbrack G_{ix},G_{jx^{\prime }}] &=&-G_{ix^{\prime }}\,\partial _{jx}\delta
(x,x^{\prime })-G_{jx}\,\partial _{ix}\delta (x,x^{\prime })~.  \label{LB3}
\end{eqnarray}%
The previous quotes in \textquotedblleft group\textquotedblright\ and
\textquotedblleft algebra\textquotedblright\ are a reminder that strictly,
the set of deformations do not form a group. The composition of two
successive deformations is itself a deformation, of course, but it also
depends on the surface to which the first deformation is applied. Thus, the
\textquotedblleft structure constants\textquotedblright\ on the right hand
sides of (\ref{LB1}-\ref{LB3}) are not constant, they depend on the surface $%
\sigma $ through its metric $g_{ij}$ which appears explicitly on the right
hand side of (\ref{LB1}).

\section{Consistent entropic dynamics: path independence}

To obtain a fully covariant dynamics we require that the evolution of any
dynamical quantity such as $\rho _{\sigma }[\chi ]$ from an initial $\sigma
_{i}$ to a final $\sigma _{f}$ be consistent with the kinematics of surface
deformations. Thus, the requirement of embeddability translates into a
consistency requirement of \emph{path independence}: if there are different
paths to evolve from an initial instant into a final instant, then all these
paths must lead to the same final values for all quantities.

In ED the relevant physical information --- supplied through the prior (\ref%
{prior}) and the constraint (\ref{Constraint 2}) --- have led us to a
diffusive\ dynamics in which the probability $\rho _{\sigma }[\chi ]$
evolves under the action of the externally prescribed drift potential $\phi
\lbrack \chi ]$. It is a curious diffusion in a curved background
space-time, but it is a diffusion nonetheless. This, however, is not what we
seek: it is not a \emph{quantum} dynamics.

A quantum dynamics requires a different choice of constraints. Specifically,
in the ED developed in the previous sections there is one basic dynamical
variable, the distribution $\rho _{\sigma }[\chi ]$. The drift potential $%
\phi \lbrack \chi ]$, being externally prescribed, is not a dynamical
variable. In contrast, in a quantum dynamics there are two dynamical
variables, the magnitude and the phase of the wave function. An additional
degree of freedom must be introduced into ED. Perhaps the simplest way is to
replace the fixed prescribed potential $\phi \lbrack \chi ]$ in constraint (%
\ref{Constraint 2}) by an evolving drift potential $\phi _{\sigma }[\chi ]$
that is updated at every step in local time in response to the evolving $%
\rho _{\sigma }[\chi ]$. In such an ED every infinitesimal step in local
time involves two updates: one is the entropic update of $\rho _{\sigma
}[\chi ]$, the other is updating the constraint represented by $\phi
_{\sigma }[\chi ]$. The resulting ED describes the coupled evolution of $%
\rho _{\sigma }[\chi ]$ and $\phi _{\sigma }[\chi ].$

The obvious question is how should the potential $\phi _{\sigma }[\chi ]$ be
updated? The LTFP equations, particularly when written in the form (\ref{FP
equation H}), suggest that the rules for updating $\phi _{\sigma }$ are more
conveniently expressed in terms of the transformed variable $\Phi _{\sigma }$%
. In previous work on ED \cite{Caticha 2010}\cite{Bartolomeo et al 2014}\cite%
{Ipek Caticha 2014} the basic criterion involved imposing the conservation
of a global energy functional. This \emph{non-dissipative} diffusion led to
a fully Hamiltonian formalism with $\rho _{\sigma }[\chi ]$ and $\Phi
_{\sigma }[\chi ]$ as conjugate variables: one Hamilton equation describes
the entropic evolution of $\rho _{\sigma }[\chi ]$, while the conjugate
Hamilton equation describes the evolving constraints through $\Phi _{\sigma
}[\chi ]$. In a covariant ED involving local surface deformations this is
not satisfactory because the notion of a global energy is not available.
Here we propose instead that the update of $\Phi _{\sigma }$ must reflect
the kinematics of surface deformations. We require path independence; the
update must be independent of the selected foliation. Next we tackle the
problem of implementing this proposal.

We saw that in an inference-based framework such as ED the concept of time
is \emph{designed} so that each instant --- which includes a specification
of both the surface $\sigma $ and the statistical state $\rho _{\sigma
}[\chi ]$ and $\Phi _{\sigma }[\chi ]$ --- contains the relevant information
to construct the next instant. This means that by design in ED time is
constructed so that \textquotedblleft given the present, the future is
independent of the past.\textquotedblright\ Thus in ED the equations of
motion will necessarily be first order in time. We have also seen that in
the limit of flat space limit --- whether relativistic \cite{Ipek Caticha
2014}, or not \cite{Bartolomeo et al 2014}\cite{Caticha 2017} --- the
non-dissipative ED is Hamiltonian. Therefore it is natural to adopt a
Hamiltonian formalism in which $\Phi _{\sigma }$ is the momentum canonically
conjugate to $\rho _{\sigma }$.

The assumption of an underlying symplectic structure is a strong one that
demands justification. In the context of non-relativistic quantum mechanics
the symplectic and complex structures characteristic of quantum mechanics
can be motivated using arguments from information geometry \cite{Caticha
2017}.

Once a Hamiltonian framework is adopted, we can follow Dirac and treat the
surface variables as if they were dynamical variables too. This allows a
Hamiltonian formalism that treats dynamical and kinematical variables in a
unified way. To do this one formally introduces auxiliary variables $\pi
_{\mu }(x)=\pi _{\mu x}$ that will play the role of momenta conjugate to $%
X_{x}^{\mu }$. These $\pi $'s are defined through the Poisson bracket (PB)\
relations, 
\begin{equation}
\lbrack X_{x}^{\mu },X_{x^{\prime }}^{\nu }]=0\,,\quad \lbrack \pi _{\mu
x},\pi _{\nu x^{\prime }}]=0\,,\quad \lbrack X_{x}^{\mu },\pi _{\nu
x^{\prime }}]=\delta _{\nu }^{\mu }\delta (x,x^{\prime })~.  \label{PB HS}
\end{equation}%
The canonical pairs $(X_{x}^{\mu },\pi _{\mu x})$ represent the geometry of
the surfaces and how they change along the foliation.

The change of a generic functional $T[X,\pi ,\rho ,\Phi ]$ resulting from an
arbitrary deformation $\delta \xi _{x}^{A}=(\delta \xi _{x}^{\bot },\delta
\xi _{x}^{i})$ is expressed in terms of PBs, 
\begin{equation}
\delta T=\int dx\,\delta \xi _{x}^{\mu }\left[ T,H_{\mu x}\right] =\int
dx\,\left( \delta \xi _{x}^{\bot }\left[ T,H_{\bot x}\right] +\delta \xi
_{x}^{i}\left[ T,H_{ix}\right] \right) ~,  \label{PB evol eq}
\end{equation}%
where $H_{\bot x}[X,\pi ,\rho ,\Phi ]$ and $H_{ix}[X,\pi ,\rho ,\Phi ]$ are
the generators of normal and tangential deformations respectively, and the
generic PB of two arbitrary functionals $U$ and $V$ is%
\begin{equation}
\left[ U,V\right] =\int dx\,\left( \frac{\delta U}{\delta X_{x}^{\mu }}\frac{%
\delta V}{\delta \pi _{\mu x}}-\frac{\delta U}{\delta \pi _{\mu x}}\frac{%
\delta V}{\delta X_{x}^{\mu }}\right) +\int D\chi \,\left( \frac{\tilde{%
\delta}U}{\tilde{\delta}\rho }\frac{\tilde{\delta}V}{\tilde{\delta}\Phi }-%
\frac{\tilde{\delta}U}{\tilde{\delta}\Phi }\frac{\tilde{\delta}V}{\tilde{%
\delta}\rho }\right) .  \label{General PB}
\end{equation}%
Thus, the PBs perform a double duty: on one hand they reflect the kinematics
of deformations of surfaces embedded in a background space-time, and on the
other hand they express the genuine entropic dynamics of $\rho $ and $\Phi $.

To comply with the requirement of path independence we follow Teitelboim and
Kucha\u{r} \cite{Teitelboim 1973a}-\cite{Kuchar 1973} to seek generators $%
H_{\bot x}$ and $H_{ix}$ that provide a \emph{canonical} \emph{representation%
} of the DHKT \textquotedblleft algebra\textquotedblright\ of surface
deformations. Unlike DHKT who developed a classical formalism based on
choosing the field $\chi (x)$ and its momentum as canonical variables, here
we develop a quantum formalism. We choose the functionals $\rho \lbrack \chi
]$ and $\Phi \lbrack \chi ]$ as the pair of canonical variables.

The idea then is that in order for the dynamics to be consistent with the
kinematics of surface deformations the PBs of $H_{\bot x}$ and $H_{ix}$ must
close in the same way as the \textquotedblleft group\textquotedblright\ of
deformations (\ref{LB1}-\ref{LB3}) --- that is, they must provide a
\textquotedblleft representation\textquotedblright\ involving the same
\textquotedblleft structure constants\textquotedblright ,\footnote{%
The difference in sign in the Poisson brackets (\ref{PB 1}-\ref{PB 3})
relative to the Lie brackets (\ref{LB1}-\ref{LB3}) arises from the change $%
\delta T$ in (\ref{PB evol eq}) being written in terms of $\left[ T,H_{\mu x}%
\right] $ rather than $\left[ H_{\mu x},T\right] $.} 
\begin{eqnarray}
\lbrack H_{\bot x},H_{\bot x^{\prime }}] &=&(g_{x}^{ij}H_{jx}+g_{x^{\prime
}}^{ij}H_{jx^{\prime }})\partial _{ix}\delta (x,x^{\prime })~,  \label{PB 1}
\\
\lbrack H_{ix},H_{\bot x^{\prime }}] &=&H_{\bot x}\partial _{ix}\delta
(x,x^{\prime })~,  \label{PB 2} \\
\lbrack H_{ix},H_{jx^{\prime }}] &=&H_{ix^{\prime }}\,\partial _{jx}\delta
(x,x^{\prime })+H_{jx}\,\partial _{ix}\delta (x,x^{\prime })~.  \label{PB 3}
\end{eqnarray}%
It may be worth noting that these equations have not been derived; it is
more appropriate to say that imposing (\ref{PB 1}-\ref{PB 3}) as strong
constraints constitutes our \emph{definition} of what we mean by a
\textquotedblleft representation\textquotedblright . To complete the
definition, we add that, as shown in \cite{Teitelboim 1973a}\cite{Teitelboim
1973b}, the requirement that the evolution of an arbitrary functional $%
T[X,\pi ,\rho ,\Phi ]$ satisfy path independence implies that the initial
values of the canonical variables must be restricted to obey the weak
constraints 
\begin{equation}
H_{\bot x}\approx 0\quad \text{and}\quad H_{ix}\approx 0~.
\label{Hp Hi constr}
\end{equation}%
Furthermore, once satisfied on an initial surface $\sigma $ the dynamics
will be such as to preserve (\ref{Hp Hi constr}) for all subsequent surfaces
of the foliation.

\section{The canonical representation}

Next we seek explicit expressions for $H_{\bot x}$ and $H_{ix}$. A surface
deformation is described by (\ref{deformation vector}), 
\begin{equation}
\delta X_{x}^{\mu }=\delta \xi _{x}^{\bot }n_{x}^{\mu }+\delta \xi
_{x}^{i}X_{ix}^{\mu }~.  \label{HS deform a}
\end{equation}%
On the other hand, we can evaluate $\delta X_{x}^{\mu }$ using (\ref{General
PB}), 
\begin{equation}
\delta X_{x}^{\mu }=\int dx^{\prime }\left( [X_{x}^{\mu },H_{\bot x^{\prime
}}]\delta \xi _{x^{\prime }}^{\bot }+[X_{x}^{\mu },H_{ix^{\prime }}]\delta
\xi _{x^{\prime }}^{i}\right) ~.  \label{HS deform b}
\end{equation}%
Since 
\begin{equation}
\lbrack X_{x}^{\mu },H_{\bot x^{\prime }}]=\frac{\delta H_{\bot x^{\prime }}%
}{\delta \pi _{\mu x}}\quad \text{and}\quad \lbrack X_{x}^{\mu
},H_{ix^{\prime }}]=\frac{\delta H_{ix^{\prime }}}{\delta \pi _{\mu x}}~,
\end{equation}%
comparing (\ref{HS deform a}) and (\ref{HS deform b}) leads to 
\begin{equation}
\frac{\delta H_{\bot x^{\prime }}}{\delta \pi _{\mu x}}=n_{x}^{\mu }\,\delta
(x,x^{\prime })\quad \text{and}\quad \frac{\delta H_{ix^{\prime }}}{\delta
\pi _{\mu x}}=X_{ix}^{\mu }\,\delta (x,x^{\prime })~.
\end{equation}%
These equations can be integrated to give, 
\begin{equation}
H_{\bot x}=\pi _{\bot x}+\tilde{H}_{\bot }\quad \text{and}\quad H_{ix}=\pi
_{ix}+\tilde{H}_{ix}~,  \label{gen a}
\end{equation}%
where 
\begin{equation}
\pi _{\bot x}=n_{x}^{\mu }\,\pi _{\mu x}\quad \text{and}\quad \pi
_{ix}=X_{ix}^{\mu }\pi _{\mu x}~,  \label{gen b}
\end{equation}%
and $\tilde{H}_{\bot }$ and $\tilde{H}_{i}$ are constants of integration
that are independent of the surface momenta $\pi _{\mu }$ but can in
principle depend on the other canonical variables, $X$, $\rho $, and $\Phi $.

Thus, the generators $H_{\perp x}$ and $H_{ix}$ separate into two
components: one pair, $\pi _{\bot x}$ and $\pi _{ix}$, that acts only on the
geometry and another pair, $\tilde{H}_{\bot x}$ and $\tilde{H}_{ix}$,\ that
acts both on the matter variables\footnote{%
We call `matter' any quantity that is not `geometry'. It is an abuse of
language to refer to the epistemic quantities $\rho $ and $\Phi $ as
`matter' but it is nevertheless convenient to do so.} and the geometry. The
latter, $\tilde{H}_{\bot x}$ and $\tilde{H}_{ix}$, are called the ensemble
Hamiltonian and the ensemble momentum. In what follows these will be
abbreviated to e-Hamiltonian and e-momentum respectively.

It is a lengthy but straightforward algebraic exercise to check that $\pi
_{\bot x}$ and $\pi _{ix}$ satisfy the DHKT \textquotedblleft
algebra\textquotedblright , eqs.(\ref{PB 1}-\ref{PB 3}), 
\begin{eqnarray}
\lbrack \pi _{\bot x},\pi _{\bot x^{\prime }}] &=&(g_{x}^{ij}\pi
_{jx}+g_{x^{\prime }}^{ij}\pi _{jx^{\prime }})\partial _{ix}\delta
(x,x^{\prime })~,  \label{pi PB 1} \\
\lbrack \pi _{ix},\pi _{\bot x^{\prime }}] &=&\pi _{\bot x}\partial
_{ix}\delta (x,x^{\prime })~,  \label{pi PB 2} \\
\lbrack \pi _{ix},\pi _{jx^{\prime }}] &=&\pi _{ix^{\prime }}\,\partial
_{jx}\delta (x,x^{\prime })+\pi _{jx}\,\partial _{ix}\delta (x,x^{\prime })~.
\label{pi PB 3}
\end{eqnarray}

\subsection{The e-momentum generators}

The generators of tangential deformations are the simpler ones: they induce
translations of the dynamical variables parallel to the surface. The change
in $\rho $ and $\Phi $ (and functionals thereof) under a tangential
deformation $\delta \xi _{x}^{i}$ is\footnote{%
We are comparing new $\chi ^{\prime }$ and old $\chi $ fields located at
different points with the same coordinate $x$. Under a tangential
deformation $\delta \xi ^{a}$ the new field is\ $\chi ^{\prime }(x)=\chi
(x+\delta \xi )$ so that 
\begin{equation*}
\delta \chi _{x}=\chi (x+\delta \chi )-\chi (x)=(\partial _{ix}\chi
_{x})\delta \xi _{x}^{i}~.
\end{equation*}%
}

\begin{equation}
\frac{\delta \rho }{\delta \xi _{x}^{i}}=\frac{\delta \rho }{\delta \chi _{x}%
}(\partial _{ix}\chi _{x})\quad \text{and}\quad \frac{\delta \Phi }{\delta
\xi _{x}^{i}}=\frac{\delta \Phi }{\delta \chi _{x}}(\partial _{ix}\chi
_{x})~.  \label{tang deformation}
\end{equation}%
This change is generated by the e-momentum $\tilde{H}_{ix}$ according to

\begin{eqnarray}
\frac{\delta \rho }{\delta \xi _{x}^{i}} &=&[\rho ,\tilde{H}_{ix}]=\frac{%
\tilde{\delta}\tilde{H}_{ix}}{\tilde{\delta}\Phi }~,  \label{tang def b} \\
\frac{\delta \Phi }{\delta \xi _{x}^{i}} &=&[\Phi ,\tilde{H}_{ix}]=-\frac{%
\tilde{\delta}\tilde{H}_{ix}}{\tilde{\delta}\rho }~.  \label{tang def c}
\end{eqnarray}%
One can easily check that the required e-momentum is 
\begin{equation}
\tilde{H}_{ix}[\rho ,\Phi ]=-\int D\chi \,\rho \lbrack \chi ]\frac{\delta
\Phi \lbrack \chi ]}{\delta \chi _{x}}\,\partial _{ix}\chi _{x}~,
\label{e-Momentum}
\end{equation}%
which shows that $\tilde{H}_{ix}[\rho ,\Phi ;\chi ]$ has an explicit
dependence on $\chi _{x}$ but is independent of the surface variables $%
X^{\mu }$. It is also straightforward to check that $\tilde{H}_{ix}$
satisfies the condition (\ref{PB 3}),%
\begin{equation}
\lbrack \tilde{H}_{ix},\tilde{H}_{jx^{\prime }}]=\tilde{H}_{ix^{\prime
}}\,\partial _{jx}\delta (x,x^{\prime })+\tilde{H}_{jx}\,\partial
_{ix}\delta (x,x^{\prime })~,  \label{e-H PB 3}
\end{equation}%
so that the tangential generators $\pi _{ix}$ and $\tilde{H}_{ix}$ satisfy (%
\ref{PB 3}) separately.

\subsection{The e-Hamiltonian generators}

The mixed PB relations, eq.(\ref{PB 2}), are the easiest to satisfy and
therefore the least informative. Using (\ref{pi PB 2}) and $[\tilde{H}%
_{ix},\pi _{\bot x^{\prime }}]=0$ we have 
\begin{equation}
\lbrack \pi _{ix}+\tilde{H}_{ix},\tilde{H}_{\bot x^{\prime }}]=\tilde{H}%
_{\bot x}\partial _{ix}\delta (x,x^{\prime })~,
\end{equation}%
which tells us that $\tilde{H}_{\bot x}$ is a scalar density. In contrast,
the normal PB relations, eq.(\ref{PB 1}) which we use (\ref{pi PB 1}) to
re-write as 
\begin{equation}
\lbrack \tilde{H}_{\bot x},\tilde{H}_{\bot x^{\prime }}]+[\pi _{\bot x},%
\tilde{H}_{\bot x^{\prime }}]+[\tilde{H}_{\bot x},\pi _{\bot x^{\prime
}}]=(g_{x}^{ij}\tilde{H}_{jx}+g_{x^{\prime }}^{ij}\tilde{H}_{jx^{\prime
}})\partial _{ix}\delta (x,x^{\prime })~,  \label{e-H PB 1a}
\end{equation}%
are crucial. They provide the criteria for updating the constraints that
define the entropic dynamics. Thus, our goal is to find a functional $\tilde{%
H}_{\bot x}[X,\rho ,\Phi ;\chi ]$ that generates a path-independent entropic
dynamics --- that is, it reproduces the local time Fokker-Planck equations (%
\ref{FP equation}), while remaining consistent with the algebra of
deformations.

The desired $\tilde{H}_{\bot x}$ is given by (\ref{e-Hp a}), and the
relation (\ref{e-H PB 1}) will serve to determine the so-far unknown
e-functional $F_{x}[X,\rho ;\chi ]$. Finding the most general solution of (%
\ref{e-H PB 1a}) lies beyond the scope of this paper; what we will do is to
identify a sufficiently large class of solutions that proves to be of
physical interest.

We will restrict our search to situations where $\tilde{H}_{\bot x}$ (and $%
F_{x}$) depend on the geometric variables $X_{x}^{\mu }$ only through the
metric $g_{ijx}$ and not through any of its derivatives. Then, using 
\begin{equation}
\frac{\delta g_{ijx^{\prime }}}{\delta \xi _{x}^{\bot }}=n_{x}^{\mu }\frac{%
\delta g_{ijx^{\prime }}}{\delta X_{x}^{\mu }}=2K_{ijx}\,\delta (x^{\prime
},x)
\end{equation}%
where $K_{ijx}$ is the extrinsic curvature, we find that 
\begin{equation}
\lbrack \pi _{\bot x},\tilde{H}_{\bot x^{\prime }}]=2K_{ijx}\,\frac{\partial 
\tilde{H}_{\bot x}}{\partial g_{ijx}}\delta (x^{\prime },x)
\end{equation}%
is symmetric in $(x^{\prime },x)$. Therefore 
\begin{equation}
\lbrack \pi _{\bot x},\tilde{H}_{\bot x^{\prime }}]+[\tilde{H}_{\bot x},\pi
_{\bot x^{\prime }}]=0~,
\end{equation}%
and (\ref{e-H PB 1a}) simplifies to

\begin{equation}
\lbrack \tilde{H}_{\bot x},\tilde{H}_{\bot x^{\prime }}]=(g_{x}^{ij}\tilde{H}%
_{jx}+g_{x^{\prime }}^{ij}\tilde{H}_{jx^{\prime }})\partial _{ix}\delta
(x,x^{\prime })~.  \label{e-H PB 1}
\end{equation}%
Thus, the normal generators $\pi _{\bot x}$ and $\tilde{H}_{\bot x}$ satisfy
(\ref{PB 1}) separately \cite{Teitelboim 1973b}.

Next we turn out attention to finding a class of physically interesting
e-functionals $F_{x}[X,\rho ;\chi ]$. We proceed in steps. First we re-write
(\ref{e-Hp a}) in the form,%
\begin{equation}
\tilde{H}_{\perp x}[X,\rho ,\Phi ;\chi ]=\tilde{H}_{\perp x}^{0}[X,\rho
,\Phi ;\chi ]+F_{x}^{0}\left[ X,\rho ;\chi \right]  \label{e-Hp b}
\end{equation}%
where%
\begin{equation}
\tilde{H}_{\perp x}^{0}=\int D\chi \left\{ \frac{1}{2}\frac{\rho }{%
g_{x}^{1/2}}\left( \frac{\delta \Phi }{\delta \chi _{x}}\right) ^{2}+\rho 
\frac{g_{x}^{1/2}}{2}g^{ij}\partial _{i}\chi _{x}\partial _{j}\chi
_{x}\right\} ~.  \label{e-Hp c}
\end{equation}%
This amounts to a mere definition of a new $F_{x}^{0}$ in terms of the old $%
F_{x}$ so there is no loss of generality. The reason that adding the second
term in (\ref{e-Hp c}) turns out to be convenient is that the new $\tilde{H}%
_{\perp x}^{0}$ satisfies (\ref{e-H PB 1}), 
\begin{equation}
\lbrack \tilde{H}_{\bot x}^{0},\tilde{H}_{\bot x^{\prime }}^{0}]=(g_{x}^{ij}%
\tilde{H}_{jx}+g_{x^{\prime }}^{ij}\tilde{H}_{jx^{\prime }})\partial
_{ix}\delta (x,x^{\prime })~.
\end{equation}%
Then, substituting (\ref{e-Hp b}) into (\ref{e-H PB 1}), and noting that 
\begin{equation}
\lbrack F_{x}^{0},F_{x^{\prime }}^{0}]=0~,
\end{equation}%
leads to 
\begin{equation}
\lbrack \tilde{H}_{\bot x}^{0},F_{x^{\prime }}^{0}]=[\tilde{H}_{\bot
x^{\prime }}^{0},F_{x}^{0}]~,  \label{PB of HF}
\end{equation}%
which is a homogeneous and linear equation for $F_{x}^{0}$. Thus, the
condition for a functional $F_{x}^{0}$ to be acceptable is that the PB $[%
\tilde{H}_{\bot x}^{0},F_{x^{\prime }}^{0}]$ be symmetric under the exchange
of $x$ and $x^{\prime }$.

The next step is to calculate the PB on the left, 
\begin{equation}
\lbrack \tilde{H}_{\bot x}^{0},F_{x^{\prime }}^{0}]=-\int D\chi \frac{\tilde{%
\delta}\tilde{H}_{\bot x}^{0}}{\tilde{\delta}\Phi }\frac{\tilde{\delta}%
F_{x^{\prime }}^{0}}{\tilde{\delta}\rho }\ ,  \label{PB HF a}
\end{equation}%
and note that $\tilde{\delta}\tilde{H}_{\bot x}^{0}/\tilde{\delta}\Phi $
reproduces the LTFP eq.(\ref{FP equation}) and (\ref{FP equation H}), 
\begin{equation}
\frac{\tilde{\delta}\tilde{H}_{\bot x}^{0}}{\tilde{\delta}\Phi }%
=-\,g_{x}^{-1/2}\frac{\delta }{\delta \chi _{x}}\left( \rho \,\frac{\delta
\Phi }{\delta \chi _{x}}\right) ~.  \label{PB HF b}
\end{equation}%
We restrict our search further by considering e-functionals $%
F_{x}^{0}[X,\rho ;\chi ]$ of the form, 
\begin{equation}
F_{x}^{0}[X,\rho ;\chi ]=\int D\chi \,f_{x}\left( X_{x},\rho ,\frac{\delta
\rho }{\delta \chi _{x}};\chi _{x},\partial \chi _{x}\right) ~,
\label{local F}
\end{equation}%
where $f_{x}$ is a \emph{function} (not a functional) of its arguments. For
such a special $F_{x}^{0}$ we have 
\begin{equation}
\frac{\tilde{\delta}F_{x^{\prime }}^{0}}{\tilde{\delta}\rho }=\frac{\partial
f_{x^{\prime }}}{\partial \rho }-\frac{\delta }{\delta \chi _{x^{\prime }}}%
\frac{\partial f_{x^{\prime }}}{\partial (\delta \rho /\delta \chi
_{x^{\prime }})}~.  \label{PB HF c}
\end{equation}%
Substituting (\ref{PB HF b}) and (\ref{PB HF c}) into (\ref{PB HF a}) gives 
\begin{equation}
\lbrack \tilde{H}_{\bot x}^{0},F_{x^{\prime }}^{0}]=\int D\chi \frac{1}{%
g_{x}^{1/2}}\frac{\delta }{\delta \chi _{x}}\left( \rho \,\frac{\delta \Phi 
}{\delta \chi _{x}}\right) \left( \frac{\partial f_{x^{\prime }}}{\partial
\rho }-\frac{\delta }{\delta \chi _{x^{\prime }}}\frac{\partial f_{x^{\prime
}}}{\partial (\delta \rho /\delta \chi _{x^{\prime }})}\right) \ .
\label{PB HF d}
\end{equation}%
Any $F_{x}^{0}$, whether of type (\ref{local F}) or not, must be a scalar
density which means that $f_{x}$ must be a scalar density too. Since the
available scalar densities are $g_{x}^{1/2}$ and $\delta /\delta \chi _{x}$,
some natural proposals are 
\begin{equation}
f_{x}\sim g_{x}^{1/2}\rho \chi _{x}^{n}\quad \text{(integer }n\text{)}\quad 
\text{and}\quad f_{x}\sim \frac{1}{g_{x}^{1/2}\rho }\left( \frac{\delta \rho 
}{\delta \chi _{x}}\right) ^{2}~.  \label{PB HF e}
\end{equation}%
A straightforward substitution into (\ref{PB HF d}) shows that all these
trials satisfy (\ref{PB of HF}) --- indeed, the PB $[\tilde{H}_{\bot
x}^{0},F_{x^{\prime }}^{0}]$ is symmetric in $(x,x^{\prime })$.\footnote{%
A more systematic study, carried out in forthcoming work, shows that trials
of the form 
\begin{equation*}
f_{x}\sim g_{x}^{1/2}\left( \frac{1}{g_{x}^{1/2}}\frac{\delta \rho }{\delta
\chi _{x}}\right) ^{k}\rho ^{l}\chi _{x}^{m}(\partial \chi _{x})^{n}\,,
\end{equation*}%
where $k$, $\ell $, $m$, and $n$ are integers, are ruled out except for (\ref%
{PB HF e}).} Finally, since (\ref{PB of HF}) is linear we can also consider
linear combinations of these trial forms which leads to a generic potential
describing self-interactions and interactions with the background geometry, 
\begin{equation}
V(\chi _{x},X_{x})=\sum_{n\ell }\lambda _{n}\chi _{x}^{n}~,  \label{V}
\end{equation}%
where $\lambda _{n}$ are coupling constants. We have therefore shown that
the family of Hamiltonians 
\begin{eqnarray}
\tilde{H}_{\perp x} &=&\int D\chi \rho\left\{ \frac{1}{2\, g_{x}^{1/2}}%
\left( \frac{\delta \Phi }{\delta \chi _{x}}\right) ^{2}+ \frac{g_{x}^{1/2}}{%
2}g^{ij}\partial _{i}\chi _{x}\partial _{j}\chi _{x}\right.  \notag \\
&&+\left. g_{x}^{1/2} V(\chi _{x})+\frac{\lambda }{g_{x}^{1/2} }\left( \frac{%
\delta \log\rho }{\delta \chi _{x}}\right) ^{2}\right\} ~,~  \label{e-Hp}
\end{eqnarray}%
generates a path-independent entropic dynamics. To interpret the last term
in (\ref{e-Hp}) we recall that in flat space-time the quantum potential is
given by \cite{Ipek Caticha 2014} 
\begin{equation}
Q=\int D\chi \int d^{3}x\frac{\lambda }{\rho }\left( \frac{\delta \rho }{%
\delta \chi _{x}}\right) ^{2}~.\,
\end{equation}%
The transition to curved coordinates and to curved space-time is achieved by
setting 
\begin{equation}
d^{3}x\rightarrow g_{x}^{1/2}d^{3}x\quad \text{and}\quad \frac{\delta }{%
\delta \chi _{x}}\rightarrow g_{x}^{-1/2}\frac{\delta }{\delta \chi _{x}}
\end{equation}%
which gives 
\begin{equation}
Q_{\sigma }=\int D\chi \int d^{3}x\frac{\lambda }{g_{x}^{1/2}\rho }\left( 
\frac{\delta \rho }{\delta \chi _{x}}\right) ^{2}~.\,  \label{Q}
\end{equation}%
Therefore the last term in (\ref{e-Hp}) may be called the \textquotedblleft
local quantum potential.\textquotedblright\ Its contribution to the energy
is such that those states that are more smoothly spread out in configuration
space tend to have lower energy. The corresponding coupling constant $%
\lambda >0$ controls the relative importance of the quantum potential; the
case $\lambda <0$ is excluded because it leads to instabilities.

\section{The evolution equations}

We will now summarize the main results of the previous sections by writing
down the equations that describe how the probability distribution $\rho
\lbrack \chi ]$ evolves in a curved space-time.

\subsection{Entropic Dynamics in a curved space-time}

Given a space-time with metric $g_{\mu \nu }(X)$ we start by specifying a
foliation of surfaces $\sigma _{t}$ labeled by a time parameter $t$, $X^{\mu
}=X^{\mu }(x,t)$, where $x^{i}$ are coordinates on the surface. The metric
induced on the surface is given by (\ref{induced metric}). The deformation
of $\sigma _{t}$ to $\sigma _{t+dt}$ is given by (\ref{deformation vector}), 
\begin{equation}
\delta \xi ^{\mu }=\delta \xi ^{\bot }n^{\mu }+\delta \xi ^{i}X_{i}^{\mu
}=[N_{xt}n^{\mu }+N_{xt}^{i}X_{i}^{\mu }]dt~,
\end{equation}%
where we introduced the scalar \emph{lapse}, $N_{xt}=\delta \xi _{x}^{\bot
}/dt$, and the vector \emph{shift}, $N_{xt}^{i}=\delta \xi _{x}^{i}/dt$.

The goal is to determine the evolution of the probability distribution $\rho
_{t}[\chi ]$ with time $t$. This requires finding the evolution of the phase
functional, $\Phi _{t}[\chi ]$. The evolution of $\rho _{t}$ and $\Phi _{t}$
is given by (\ref{PB evol eq}), 
\begin{equation}
\frac{\partial \rho _{t}}{\partial t}=\left[ \rho _{t},H\right] \quad \text{%
and}\quad \frac{\partial \Phi _{t}}{\partial t}=\left[ \Phi _{t},H\right] ~,
\end{equation}%
where $H$ is the smeared Hamiltonian, 
\begin{equation}
H[N,N^{i}]=\int dx\,\left( N_{xt}H_{\bot x}+N_{xt}^{i}H_{ix}\right) ~.\label{Smeared Hamiltonian}
\end{equation}%
The result is 
\begin{equation}
\frac{\partial \rho _{t}}{\partial t}=\int d^{3}x\,\left( \frac{\delta \rho
_{t}}{\delta \xi _{x}^{\bot }}N_{xt}+\frac{\delta \rho _{t}}{\delta \xi
_{x}^{i}}N_{xt}^{i}\right)
\end{equation}%
and 
\begin{equation}
\frac{\partial \Phi _{t}}{\partial t}=\int d^{3}x\,\left( \frac{\delta \Phi
_{t}}{\delta \xi _{x}^{\bot }}N_{xt}+\frac{\delta \Phi _{t}}{\delta \xi
_{x}^{i}}N_{xt}^{i}\right) ~.
\end{equation}%
The tangential derivatives, $\delta \rho _{t}/\delta \xi _{x}^{i}$ and $%
\delta \Phi _{t}/\delta \xi _{x}^{i}$, are given by eqs.(\ref{tang
deformation}-\ref{tang def c}). The normal derivatives, $\delta \rho
_{t}/\delta \xi _{x}^{\bot }$ and $\delta \Phi _{t}/\delta \xi _{x}^{\bot }$%
, are given by 
\begin{eqnarray}
\frac{\delta \rho _{t}}{\delta \xi _{x}^{\bot }} &=&[\rho _{t},\tilde{H}%
_{\bot x}]=\frac{\tilde{\delta}\tilde{H}_{\bot x}}{\tilde{\delta}\Phi _{t}}~,
\\
\frac{\delta \Phi _{t}}{\delta \xi _{x}^{\bot }} &=&[\Phi _{t},\tilde{H}%
_{\bot x}]=-\frac{\tilde{\delta}\tilde{H}_{\bot x}}{\tilde{\delta}\rho _{t}}%
~.
\end{eqnarray}%
Substituting $\tilde{H}_{\bot x}$ from (\ref{e-Hp}) gives the local-time
Fokker-Planck equations (\ref{FP equation}), 
\begin{equation}
\frac{\delta \rho _{t}}{\delta \xi _{x}^{\bot }}=-\,\frac{1}{g_{x}^{1/2}}%
\frac{\delta }{\delta \chi _{x}}\left( \rho _{t}\,\frac{\delta \Phi _{t}}{%
\delta \chi _{x}}\right) ~,  \label{LT FP}
\end{equation}%
and the local time generalization of the Hamilton-Jacobi\ equations, 
\begin{equation}
-\frac{\delta \Phi _{t}}{\delta \xi _{x}^{\bot }}=\frac{1}{2g_{x}^{1/2}}%
\left( \frac{\delta \Phi _{t}}{\delta \chi _{x}}\right) ^{2}+\frac{%
g_{x}^{1/2}}{2}g^{ij}\partial _{i}\chi _{x}\partial _{j}\chi
_{x}+g_{x}^{1/2}V(\chi _{x},X_{x})-\frac{4\lambda}{g_{x}^{1/2}\rho ^{1/2}}%
\frac{\delta ^{2}\rho _{t}^{1/2}}{\delta \chi _{x}^{2}}~.  \label{LT HJ}
\end{equation}%
The formulation of the ED of fields in curved space-time is thus completed.
However, it may not yet be obvious that this is a quantum theory; to make it
explicit is the next task.

\subsection{The local-time Schr\"{o}dinger functional equation}

The relation of the ED formalism to quantum theory can made explicit by
making a canonical transformation (often called a \textit{Madelung}
transformation) from the dynamical variables $\rho $ and $\Phi $ into a pair
of complex variables,\footnote{%
For non-relativistic particles the underlying symplectic and complex
structures and their relation to information geometry are derived in \cite%
{Caticha 2017}.} 
\begin{equation}
\Psi \lbrack \chi ]=\rho ^{1/2}e^{i\Phi}~.
\end{equation}%
The equation of evolution for the new variable $\Psi _{t}[\chi ]$ is then
given by 
\begin{equation}
\frac{\partial \Psi _{t}}{\partial t}=\int d^{3}x\,\left( \frac{\delta \Psi
_{t}}{\delta \xi _{x}^{\bot }}N_{xt}+\frac{\delta \Psi _{t}}{\delta \xi
_{x}^{i}}N_{xt}^{i}\right) ~.  \label{SchEq a}
\end{equation}%
The tangential derivative, $\delta \Psi _{t}/\delta \xi _{x}^{i}$, is
obtained from eq.(\ref{tang deformation}),%
\begin{equation}
\frac{\delta \Psi _{t}}{\delta \xi _{x}^{i}}=(\partial _{ix}\chi _{x})\frac{%
\delta \Psi _{t}}{\delta \chi _{x}}~.
\end{equation}%
Using (\ref{e-Hp}) the normal derivative, 
\begin{equation}
\frac{\delta \Psi _{t}}{\delta \xi _{x}^{\bot }}=[\Psi _{t},\tilde{H}_{\bot
x}]~,
\end{equation}%
gives the local time version of Schr\"{o}dinger equation for the wave
functional $\Psi _{t}[\chi ]$, 
\begin{equation}
i\hbar \frac{\delta \Psi _{t}}{\delta \xi _{x}^{\bot }}=-\frac{\hbar ^{2}}{%
2g_{x}^{1/2}}\frac{\delta ^{2}\Psi _{t}}{\delta \chi _{x}^{2}}+\frac{%
g_{x}^{1/2}}{2}g^{ij}\partial _{i}\chi _{x}\partial _{j}\chi _{x}\,\Psi
_{t}+g_{x}^{1/2}V\Psi _{t}~,  \label{SchEq b}
\end{equation}
where we have introduced $\hbar$ by setting $\lambda = \hbar^{2}/8$ and rescaling $\Phi_{t}\to \Phi_{t}/\hbar$, so that $\Phi_{t}$ has units of action.\footnote{Note also that this modification means that we have adjusted the units of $\chi$ so that $[\chi] = [\hbar]^{1/2}/\text{length}$.}

The limit of flat space-time is obtained setting $g_{x}^{1/2}=1$, $\delta
\xi _{x}^{\bot }=dt$, $N=1$, and $N^{i}=0$. Then eqs.(\ref{SchEq a}) and (%
\ref{SchEq b}) become the Schr\"{o}dinger functional equation, 
\begin{equation}
i\hbar \frac{\partial \Psi _{t}}{\partial t}=\int d^{3}x\,\left\{ -\frac{%
\hbar ^{2}}{2}\frac{\delta ^{2}\Psi _{t}}{\delta \chi _{x}^{2}}+\frac{1}{2}%
g^{ij}\partial _{i}\chi _{x}\partial _{j}\chi _{x}\,\Psi _{t}+V\Psi
_{t}\right\} ~.  \label{SchEq c}
\end{equation}%
This is quantum field theory in the Schr\"{o}dinger functional
representation \cite{Jackiw 1989}. From here one can proceed to introduce a
Hilbert space, operators, and the standard machinery of quantum field
theory. Eq.(\ref{SchEq c}) justifies identifying the expression $(8\lambda
)^{1/2}$ with Planck's constant $\hbar $. We thus see that the coupling $%
\lambda =\hbar ^{2}/8$ in (\ref{e-Hp}) plays a crucial role: it defines the
numerical value of $\hbar $ and sets the scale that separates quantum from
classical regimes.

\section{Some applications}
As laid out above, the ED that we have developed here is formally identical to the standard quantum field theory in the Schr\"{o}dinger functional representation. This means that predictions made on the basis of equations (\ref{SchEq a})-(\ref{SchEq b}) are identical to those obtained using the standard methods. And indeed, there have been a wide range of topics pursued within this formalism; including, studies of vacuum states in curved space-time \cite{Long Shore 1998}, research on the Hawking effect \cite{Freese et al 1985}, applications in cosmology \cite{Guven et al 1989}, and investigations into symmetries \cite{Floreanini et al 1987}\cite{Halliwell 1991}, just to name a few. In other words: the ED developed here, while being fully \emph{consistent} with all of these developments, it does not go beyond them. And, indeed, the purpose of the ED developed here is not to generate better techniques for calculation, but to put QFTCS on a firm conceptual foundation.

The ED formulation of QFTCS, nonetheless, is sufficiently different from the usual approaches that a demonstration of the framework is, in fact, warranted. We do this with two examples. One example illustrates the formal flexibility of the ED formalism, while the other demonstrates the conceptual clarity that an entropic framework supplies to QFTCS. Both insights may have important implications as ED moves beyond QFTCS and begins to incorporate dynamical gravity.
\subsection{The Ehrenfest equations in Entropic Dynamics}
We begin by obtaining the Ehrenfest equations for a quantum scalar field in ED. The derivation highlights many of the novel features of the ED approach; in particular, the utilization of Hamiltonians, Poisson brackets, and so on, rather than the conventional quantum tools.
\paragraph{Some background}
In ED, the focus of our inquiries are the field variables $\chi_{x}$ which have definite, but unknown values. Thus, in the absence of such definite information, our goal is to obtain an estimate of these values. One such estimate is provided by the \emph{expected value} of the field variables
\begin{equation}
\tilde{\chi}_{x} = \int D\chi \, \rho \, \chi_{x}~.\label{Expected chi}
\end{equation}
This, in turn, defines a functional $\tilde{\chi}_{x}$ on the ensemble phase space, which makes it amenable to treatment through the canonical formalism.

However, this expected value is not static and we wish to know how it changes in time. To determine this evolution, we carry over the formalism of the previous section and introduce a Hamiltonian $H[N,N^{i}]$, as in eq.(\ref{Smeared Hamiltonian}), adapted to a particular foliation with lapse $N$, shift $N^{i}$, and parameter $t$. Moreover, we are concerned with a quantum dynamics so we choose $H_{\perp x}$ and $H_{ix}$ in eq.(\ref{Smeared Hamiltonian}) to match that of eqns.(\ref{e-Momentum}) and (\ref{e-Hp}).

\paragraph*{Evolution of the expected values}
Since $\tilde{\chi}_{x} = \tilde{\chi}_{x}[\rho]$ is a Hamiltonian functional, its update can be obtained by taking the appropriate Poisson brackets for a suitably chosen Hamiltonian $H$. Indeed, the velocity of $\tilde{\chi}_{x}$ is given by
\begin{equation}
\partial_{t}\tilde{\chi}_{x} = \int dx^{\prime} \left (N_{x^{\prime}t}\left \{\tilde{\chi}_{x},H_{\perp x^{\prime}}\right \}+N^{i}_{x^{\prime}t}\left \{\tilde{\chi}_{x},H_{i x^{\prime}}\right \}\right )\label{Expected chi velocity a}
\end{equation}
with
\begin{equation}
\left\{\tilde{\chi}_{x},H_{\perp x^{\prime}}\right \} = \delta(x,x^{\prime})\int D\chi \, \rho \frac{1}{g_{x^{\prime}}}\frac{\delta\Phi}{\delta\chi_{x^{\prime}}} \label{Expected chi velocity perp}
\end{equation}
and
\begin{equation}
\left\{\tilde{\chi}_{x},H_{i x^{\prime}}\right \} =- \delta(x,x^{\prime})\int D\chi \, \rho\, \partial_{ix^{\prime}}\chi_{x^{\prime}}~.\label{Expected chi velocity tang}
\end{equation}
Taken together, eqns.(\ref{Expected chi velocity a})-(\ref{Expected chi velocity tang}) result in
\begin{equation}
\partial_{t}\tilde{\chi}_{x} =  \int D\chi \, \rho \left ( \frac{N}{g^{1/2}_{x}}\frac{\delta\Phi}{\delta\chi_{x}}-N^{i}\partial_{ix}\chi_{x}\right )~,\label{Expected chi velocity}
\end{equation}
which contains two contributions. The latter contribution in eq.(\ref{Expected chi velocity}) is just due to the shift, while the former is due to the flow of probability, which is characterized by the appearance of the current velocity \[v_{x} = \frac{1}{g^{1/2}_{x}}\frac{\delta\Phi}{\delta\chi_{x}}~, \] first introduced in the context of the LTFP equations (\ref{FP equation}).

Consider here a related quantity, the \emph{current momentum} and its expectation, the \emph{ensemble} current momentum,\footnote{Note that while the current velocity is a scalar valued quantity, the current \emph{momentum} is a scalar \emph{density} of weight one.}
\begin{equation}
P_{x} \equiv \frac{\delta\Phi}{\delta\chi_{x}} = g_{x}^{1/2} \, v_{x}\quad \text{and}\quad \tilde{P}_{x} = \int D\chi \, \rho\, P_{x}~,\label{Expected current momentum}
\end{equation}
respectively.\footnote{Alternatively, introduce the differential operator $\hat{P}_{x} = -i\delta/\delta\chi_{x}$. Then the ensemble current momentum translates in the conventional language to the expected value of this operator. That is, $\tilde{P}_{x} = \int \Psi^{*}\hat{P}_{x}\Psi$, using the complex functionals $(\Psi^{*},\Psi)$ introduced above.} We can now conveniently rewrite the velocity $\partial_{t}\tilde{\chi}_{x}$ in terms of the current momentum, yielding
\begin{equation}
\partial_{t}\tilde{\chi}_{x} = \frac{N}{g^{1/2}_{x}}\tilde{P}_{x}-N^{i}\left \langle \partial_{ix}\chi_{x} \right \rangle~,\label{Expected chi velocity b}
\end{equation}
where we have used the notation $\left\langle A \right \rangle = \int \rho A$ to denote expectation.

The advantage of introducing the current momentum (as opposed to the current velocity) is that the expected current momentum $\tilde{P}_{x}$, together with the expected field value $\tilde{\chi}_{x}$, satisfy the canonical Poisson bracket relations
\begin{equation}
 \left \{ \tilde{\chi}_{x}, \tilde{P}_{x^{\prime}}\right \} = \delta(x,x^{\prime})~. 
 \end{equation}
This seems to suggest that $\tilde{P}_{x}$ plays the role of a momentum \emph{conjugate} to $\tilde{\chi}_{x}$.

Indeed, let us take this hint seriously and compute the corresponding Hamilton's equations for this canonical pair. The time derivative of $\tilde{\chi}_{x}$ was provided earlier, in eq.(\ref{Expected chi velocity b}). The velocity of $\tilde{P}_{x}$, on the other hand, is given by
\begin{equation}
\partial_{t}\tilde{P}_{x} = \int dx^{\prime} \left (N_{x^{\prime}t}\left \{\tilde{P}_{x},H_{\perp x^{\prime}}\right \}+N^{i}_{x^{\prime}t}\left \{\tilde{P}_{x},H_{i x^{\prime}}\right \}\right )~.\label{Expected current momentum velocity a}
\end{equation}
To compute this we need the two Poisson brackets in eq.(\ref{Expected current momentum velocity a}). A quick calculation gives
\begin{equation}
\left\{\tilde{P}_{x},H_{\perp x^{\prime}}\right \} =-\int D\chi \, \rho\, g_{x^{\prime}}^{1/2}\left (\delta(x,x^{\prime})\frac{\partial V_{x^{\prime}}}{\partial \chi_{x^{\prime}}}+g^{ij}_{x^{\prime}}\partial_{ix^{\prime}}\chi_{x^{\prime}}\partial_{jx^{\prime}}\delta(x^{\prime},x)\right ) \label{Expected current momentum velocity perp}
\end{equation}
and
\begin{equation}
\left\{\tilde{P}_{x},H_{i x^{\prime}}\right \} =\int D\chi \, \rho \frac{\delta\Phi}{\delta\chi_{x^{\prime}}}\partial_{ix^{\prime}}\delta(x^{\prime},x)\label{Expected current momentum velocity tang}
\end{equation}
so that from eq.(\ref{Expected current momentum velocity a}) we obtain
\begin{equation}
\partial_{t}\tilde{P}_{x} =  \partial_{i}\left ( N\, g_{x}^{1/2} \, g^{ij} \partial_{j}\tilde{\chi}_{x}  \right ) - \partial_{i}\left (N^{i}\tilde{P}_{x}\right ) - N\, g_{x}^{1/2} \left\langle \frac{\partial V}{\partial\chi_{x}} \right \rangle~.\label{Expected current momentum velocity}
\end{equation}
Thus an initial assignment of $\tilde{\chi}_{x}$, $\tilde{P}_{x}$, and higher statistical moments of $\chi_{x}$, will be sufficient to determine the evolution of $\tilde{\chi}_{x}$, i.e. no further derivatives are required.
\paragraph*{Ehrenfest equations}
The equations (\ref{Expected chi velocity}) and (\ref{Expected current momentum velocity}) taken together have the character of classical field equations for some \textquotedblleft classical\textquotedblright\ field variables $\tilde{\chi}_{x}$ and $\tilde{P}_{x}$. In fact, it is not difficult to show that these are nothing but the Ehrenfest equations (see e.g.,\cite{Ballentine 1998}).

To see this, note that the velocity $\partial_{t}\tilde{\chi}_{x}$ is linear in the current momentum $\tilde{P}_{x}$. Now, invert this relation for $\tilde{P}_{x}$ in terms of the velocity $\partial_{t}\tilde{\chi}_{x}$ and substitute into eq.(\ref{Expected current momentum velocity}). The result is that $\tilde{\chi}_{x}$ evolves according to the equation
\begin{equation}
\hat{\Box}\tilde{\chi}_{x} =  \left\langle \frac{\partial V(\chi_{x})}{\partial\chi_{x}} \right \rangle~,\label{Ehrenfest relations}
\end{equation}
where
\begin{align}
\hat{\Box} = -\frac{1}{N\, g^{1/2}}\bar{\partial}_{t}\left [ \left ( \frac{g^{1/2}}{N}\right )\bar{\partial}_{t}\right ]-\frac{\partial_{i}N^{i}}{N^{2}}\bar{\partial}_{t}+\frac{1}{N\, g^{1/2}}\partial_{i}\left [  N\, g^{1/2} g^{ij}\partial_{j}  \right] \label{Wave operator}
\end{align}
is the wave operator in curved space-time in foliation-adapted coordinates \cite{Long Shore 1998}\cite{Gourgoulhon 2007}, and where we have introduced the operator $\bar{\partial}_{t} = \partial_{t}+N^{i}\partial_{i}$.


Equation (\ref{Ehrenfest relations}) comprises the Ehrenfest relations that we seek. An appealing feature of such relations is that they are \emph{exact}, and thus contain complete information about the underlying quantum dynamics. This makes the Ehrenfest relations ideal for probing the behavior of quantum fields and their deviations from classical behavior. For instance, it is not difficult to see that $\tilde{\chi}_{x}$ follows a classical evolution only when
\begin{equation}
\left\langle \frac{\partial V(\chi_{x})}{\partial\chi_{x}} \right \rangle =  \frac{\partial V(\left\langle\chi_{x}\right \rangle)}{\partial\chi_{x}}~.
\end{equation}
But this, of course, only occurs when the potential is itself quadratic in the field.

Indeed, choose for $V_{x}$ the potential $V_{x} = \frac{1}{2}m^{2}\chi_{x}^{2}$ so that $\partial V/\partial \chi_{x} = m^{2}\chi_{x}$. Equation (\ref{Ehrenfest relations}) then reduces to
\begin{equation}
\left (\hat{\Box}-m^{2}\right )\tilde{\chi}_{x} = 0~,\label{Ehrenfest equation KG}
\end{equation}
which is a classical Klein-Gordon equation in curved space-time (see e.g., \cite{Hollands Wald 2014}) for the expected field configuration. Thus $\tilde{\chi}_{x}$ follows --- exactly --- the classical equations of motion, which is precisely the content of Ehrenfest's theorem, familiar from non-relativistic quantum mechanics (see e.g., \cite{Ballentine 1998}). For potentials that are \emph{not} quadratic, however, we can expect deviations from classical behavior and it is legitimate to obtain quantum corrections via an approximation scheme.
\paragraph*{Comments}
As opposed to standard formulations of equations of this type, our derivation of the Ehrenfest relations was performed entirely within the framework of ED, using Hamiltonians, Poisson brackets, etc., rather than commutators and the standard quantum machinery.\footnote{Our approach to the Ehrenfest, or Ehrenfest-Heisenberg equations, follows closely that of Ashtekar and Schilling \cite{Ashtekar Schilling 1999}.} These methods, which are geometric in nature, are thus of more general applicability than the standard techniques, which rely heavily on the \emph{linearity} of quantum theory. Indeed, while the assumption of linearity has thus far proved quite robust, it is not immediately obvious that quantum gravity needs to follow suit (see e.g., \cite{Callender Huggett 2001}-\cite{Albers et al 2008}). In such cases the ED approach might provide a viable alternative framework.
\subsection{On divergences in Entropic Dynamics}
Some of the principal benefits of the ED approach are conceptual. For instance, a central difficulty of
any quantum field theory, one that is also shared by the ED formalism, is the
problem of infinities. The nature of the infinities is, however, very
different \cite{Ipek Caticha 2014}. To see this consider the limit of flat
space-time, i.e. $N=1$, $N^{i}=0$, $g^{1/2} = 1$. Setting $V=m^{2}\chi ^{2}/2$ in the Schr\"{o}dinger equation (%
\ref{SchEq c}) leads to the quantum theory of free real scalar fields \cite%
{Jackiw 1989} from which all the standard results can be recovered (see 
\emph{e.g.}, \cite{Long Shore 1998}). For example, in units such that $\hbar
=c=1$, the wave functional of the ground state is%
\begin{equation}
\Psi _{0}\left[ \chi \right] =\frac{1}{Z_{0}^{1/2}}e^{-iE_{0}t}\exp \left[ -%
\frac{1}{2}\int d^{3}x\int d^{3}y\,\,\chi _{x}G_{xy}\chi _{y}\right] ~,
\label{Psi vac}
\end{equation}%
where 
\begin{equation}
G_{xy}=\int \frac{d^{3}k}{(2\pi )^{3}}\omega _{k}\,e^{i\vec{k}\cdot (\vec{x}-%
\vec{y})}~,\quad \text{with\quad }\omega _{k}=(\vec{k}^{2}+m^{2})^{1/2}~,
\end{equation}%
and the energy of the ground state, 
\begin{equation}
E_{0}=\int d^{3}x\,\tilde{H}_{\bot x}[\Psi _{0}]~,  \label{E vac a}
\end{equation}%
is obtained from (\ref{e-Hp}) and (\ref{Psi vac}), 
\begin{equation}
\tilde{H}_{\bot x}[\Psi _{0}]=\frac{1}{2}\int D\chi \rho _{0}\left[ \left( 
\frac{\delta \Phi _{0}}{\delta \chi _{x}}\right) ^{2}+g^{ij}\partial
_{i}\chi _{x}\partial _{j}\chi _{x}+m^{2}\chi _{x}^{2}+\left( \frac{1 }{%
2\rho _{0}}\frac{\delta \rho _{0}}{\delta \chi _{x}}\right) ^{2}\right] ~.
\label{E vac b}
\end{equation}%
The result,

\begin{equation}
E_{0}=\frac{1}{2}\int d^{3}x\,G_{xx}=\int d^{3}x\int \frac{d^{3}k}{\left(
2\pi \right) ^{3}}\frac{1}{2}\omega _{k}~,  \label{E vac c}
\end{equation}%
is both infrared and ultraviolet divergent. Similarly, the expected value of
the field at any point $\vec{x}$ vanishes but its variance diverges,%
\begin{equation}
\left\langle \chi _{x}\right\rangle =0\quad \text{and}\quad \text{Var}\left[
\chi _{x}\right] =\langle \chi _{x}^{2}\rangle _{0}=\int \frac{d^{3}k}{%
\left( 2\pi \right) ^{3}}\frac{1}{2\omega _{k}}~.  \label{varchi vac}
\end{equation}%
The point we want to stress in repeating these well-known results is that in
the ED framework the divergent quantities in (\ref{E vac c}) and (\ref%
{varchi vac}) are not physical, ontic quantities. Both the e-Hamiltonian $%
\tilde{H}_{\bot x}[\Psi _{0}]$ in (\ref{E vac b}) and the variance in (\ref%
{varchi vac}) are \emph{expected} values. \emph{The infinities are not real;
they are epistemic.}

ED recognizes the role of incomplete information: the interpretation of the
diverging Var$\left[ \chi _{x}\right] $ is not that the field $\chi _{x}$
undergoes fluctuations of infinite magnitude, but rather that with the
information that is available to us we are completely unable to predict the
value of the field $\chi _{x}$ at the sharply localized point $\vec{x}$. The
infinities are epistemic: what diverges are not physical quantities but our
uncertainty about them. However, this does not mean the theory is useless.
It may be incapable of predicting some quantities but it can provide useful
predictions for many others. For example, the equal time correlations
between two field variables at different locations are perfectly finite, 
\begin{equation}
\left\langle \chi _{x}\chi _{y}\right\rangle _{0}=\int \frac{d^{3}k}{\left(
2\pi \right) ^{3}}\frac{e^{i\vec{k}\cdot \left( \vec{x}-\vec{y}\right) }}{%
2\omega _{k}}=\frac{m}{4\pi ^{2}\left\vert \vec{x}-\vec{y}\right\vert }%
K_{1}\left( m\left\vert \vec{x}-\vec{y}\right\vert \right) ~,
\end{equation}%
where $K_{1}$ is a modified Bessel function \cite{Long Shore 1998}.

\section{Discussion}

Entropic dynamics provides an inferential alternative to the standard
methods of quantization. The ED approach avoids a representation of fields as operators and any reference to the Hilbert or Fock spaces on which they presumably act. In effect this eliminates the issue of choosing among many inequivalent representations. Consequently, the operator ordering ambiguities that are characteristic of conventional quantization methods are avoided too. Indeed, many of the problems associated with the Dirac quantization method \cite{Dirac Lectures} and the laborious techniques necessary to implement it (such as the identification and elimination of second-class constraints etc.) are completely sidestepped.

The ED approach to QFTCS, however, also offers new insights to problems such as the Unruh effect at the interface between QFTCS and the quantum measurement problem.\footnote{Being an inference theory, ED is particularly well suited to tackling the quantum problem of measurement \cite{Johnson Caticha 2011}\cite{Vanslette Caticha 2016}.} Indeed, the fact that in the ED approach fields are physical
entities which at all times have definite but possibly unknown values, while \textquotedblleft particles are whatever particle detectors detect,\textquotedblright\ immediately raises the question of what is a particle within the ED framework.

Moreover, the ED approach to QFTCS leads to a theory that is Hamiltonian in character, thus retaining the powerful tools and intuitive appeal of the classical Hamiltonian framework\footnote{%
That quantum theory can be formulated as a Hamiltonian theory has been explored by many \cite{Reginatto Hall 2011}\cite{Reginatto Hall 2012}\cite{Ashtekar Schilling 1999}\cite{Kibble 1979}\cite{Heslot 1985}. The connection between the ED framework and information geometry with the symplectic/Hamiltonian structure is explored in \cite{Caticha 2017}.} but now in the context of a fully quantum theory. This is particularly attractive for a couple reasons. First is that the ED approach proceeds without ever invoking the use of linearity or of Hilbert spaces, thus leaving open the possibility that deeper theories will introduce non-linearities --- something the standard approaches cannot readily account for. Another is that the scheme introduced here allows us to borrow the methods of DHKT \cite{Hojman et al 1976}, which were used to develop \textit{classical} covariant Hamiltonian theories, but instead apply them to a theory that is inherently statistical and quantum.

This is not insignificant. A primary difficulty in formulating a theory of quantum gravity is that general relativity and quantum field theory are couched in completely different formal languages. The ED that we have developed here, however, actually helps to bridge this divide. Indeed, while a common approach to quantum gravity is to incorporate gravitation into the linear, algebraic framework of quantum theory, ED opens the door to an alternative approach wherein quantum theory more resembles general relativity. The key is to recognize the central role played by the Hamiltonian formalism, not only in the ED formulation of QFTCS here, but also in the geometrodynamics approach to general relativity. Therefore, in contrast to many other approaches to QFTCS, ED seems to offer the possibility of extending the scheme developed here towards deriving, from \emph{first principles}, a fully dynamical theory of quantum fields interacting with classical gravity \cite{Ipek Caticha 2019}.

\subsubsection*{Acknowledgements}

We would like to thank D. Bartolomeo, N. Carrara, N. Caticha, S. DiFranzo,
K. Knuth, P. Pessoa, and K. Vanslette for valuable discussions on entropic
dynamics.

\appendix\numberwithin{equation}{section}

\section{The local-time Fokker-Planck equations}

\label{appendix FP}To rewrite the dynamical equation (\ref{Evolution
equation}) in differential form consider the probability $P[\chi ,\sigma
|\chi _{0},\sigma _{0}]$ of a \emph{finite} transition from a field
configuration $\chi _{0}$ at some early surface $\sigma _{0}$ to a
configuration $\chi $ at a later $\sigma $. The result of a further
evolution from $\sigma $ to a neighboring $\sigma ^{\prime }$ obtained from $%
\sigma $ by an infinitesimal normal deformation $\delta \xi _{x}^{\bot }$ is
given by (\ref{Evolution equation}), 
\begin{equation}
P\left[ \chi ^{\prime },\sigma ^{\prime }|\chi _{0},\sigma _{0}\right] =\int
D\chi \,P\left[ \chi ^{\prime },\sigma ^{\prime }|\chi ,\sigma \right] P%
\left[ \chi ,\sigma |\chi _{0},\sigma _{0}\right] ~.
\end{equation}%
To obtain a differential equation one cannot just Taylor expand as $\delta
\xi _{x}^{\bot }\rightarrow 0$ because $P\left[ \chi ^{\prime },\sigma
^{\prime }|\chi ,\sigma \right] $ becomes a very singular object --- a delta
functional. Instead, we multiply by an arbitrary smooth test functional $T%
\left[ \chi ^{\prime }\right] $ and integrate 
\begin{eqnarray}
&&\int D\chi ^{\prime }\,P\left[ \chi ^{\prime },\sigma ^{\prime }|\chi
_{0},\sigma _{0}\right] T\left[ \chi ^{\prime }\right]  \notag \\
&=&\int D\chi \,P\left[ \chi ,\sigma |\chi _{0},\sigma _{0}\right] \int
D\chi ^{\prime }\,T\left[ \chi ^{\prime }\right] \,P\left[ \chi ^{\prime
},\sigma ^{\prime }|\chi ,\sigma \right] \,.  \label{Test Function a}
\end{eqnarray}%
Next expand the test function $T\left[ \chi ^{\prime }\right] =T[\chi
+\Delta \chi ]$ in powers of $\Delta \chi =\chi ^{\prime }-\chi $. Since $%
\chi $ is Brownian to obtain $T\left[ \chi ^{\prime }\right] $ to first
order in $\delta \xi _{x}^{\bot }$ we need to keep second order in $\Delta
\chi _{x}$, 
\begin{eqnarray}
T\left[ \chi ^{\prime }\right] &=&T\left[ \chi \right] +\int dx\frac{\delta T%
\left[ \chi \right] }{\delta \chi _{x}}\Delta \chi _{x}  \notag \\
&&+\frac{1}{2}\int dx\,dx^{\prime }\frac{\delta ^{2}T\left[ \chi \right] }{%
\delta \chi _{x}\delta \chi _{x^{\prime }}}\Delta \chi _{x}\Delta \chi
_{x^{\prime }}+\cdots .  \label{Test Function b}
\end{eqnarray}%
Use this expansion together with (\ref{Exp Step 1}) and (\ref{Fluctuations})
to obtain 
\begin{eqnarray*}
&&\int D\chi ^{\prime }\,T\left[ \chi ^{\prime }\right] \,P\left[ \chi
^{\prime },\sigma ^{\prime }|\chi ,\sigma \right] \\
&=&T\left[ \chi \right] +\int dx\,\frac{\eta \delta \xi _{x}^{\bot }}{%
g_{x}^{1/2}}\left\{ \frac{\delta T\left[ \chi \right] }{\delta \chi _{x}}\,%
\frac{\delta \phi \left[ \chi \right] }{\delta \chi _{x}}+\frac{1}{2}\frac{%
\delta ^{2}T\left[ \chi \right] }{\delta \chi _{x}^{2}}\right\} \,.
\end{eqnarray*}%
Substituting back into eq.(\ref{Test Function a}), leads to 
\begin{eqnarray}
&&\int D\chi \,\left\{ P\left[ \chi ,\sigma ^{\prime }|\chi _{0},\sigma _{0}%
\right] -P\left[ \chi ,\sigma |\chi _{0},\sigma _{0}\right] \right\} T\left[
\chi \right]  \notag \\
&=&\int dx\,\frac{\eta \delta \xi _{x}^{\bot }}{g_{x}^{1/2}}\int D\chi P%
\left[ \chi ,\sigma |\chi _{0},\sigma _{0}\right] \left\{ \frac{\delta T%
\left[ \chi \right] }{\delta \chi _{x}}\,\frac{\delta \phi \left[ \chi %
\right] }{\delta \chi _{x}}+\frac{1}{2}\frac{\delta ^{2}T\left[ \chi \right] 
}{\delta \chi _{x}^{2}}\right\}  \label{Test Function c}
\end{eqnarray}%
Since $T[\chi ]$ is arbitrary, after some integrations by parts we get 
\begin{eqnarray}
&&P\left[ \chi ,\sigma ^{\prime }|\chi _{0},\sigma _{0}\right] -P\left[ \chi
,\sigma |\chi _{0},\sigma _{0}\right] =\int dx\frac{\delta P\left[ \chi
,\sigma |\chi _{0},t_{0}\right] }{\delta \xi _{x}^{\bot }}\delta \xi
_{x}^{\bot }  \notag \\
&=&\int dx\,\frac{\eta \delta \xi _{x}^{\bot }}{g_{x}^{1/2}}\left\{ -\frac{%
\delta }{\delta \chi _{x}}\left( \,P\left[ \chi ,\sigma |\chi _{0},\sigma
_{0}\right] \frac{\delta \phi \left[ \chi \right] }{\delta \chi _{x}}\right)
+\frac{1}{2}\frac{\delta ^{2}}{\delta \chi _{x}^{2}}P\left[ \chi ,\sigma
|\chi _{0},\sigma _{0}\right] \right\} ~.  \notag \\
&&  \label{Evol eq b}
\end{eqnarray}%
Finally, for a finite evolution from $\sigma _{0}$ to $\sigma $, (\ref%
{Evolution equation}) reads, 
\begin{equation}
\rho _{\sigma }\left[ \chi \right] =\int D\chi _{0}\,P\left[ \chi ,\sigma
|\chi _{0},\sigma _{0}\right] \rho _{\sigma _{0}}\left[ \chi _{0}\right] ~.
\end{equation}%
A further infinitesimal normal deformation $\sigma \rightarrow \sigma
^{\prime }$ by $\delta \xi _{x}^{\bot }$ gives 
\begin{eqnarray}
\rho _{\sigma ^{\prime }}\left[ \chi \right] -\rho _{\sigma }\left[ \chi %
\right] &=&\int dx\frac{\delta \rho _{\sigma }\left[ \chi \right] }{\delta
\xi _{x}^{\bot }}\delta \xi _{x}^{\bot }  \notag \\
&=&\int D\chi _{0}\,\left( \int dx\frac{\delta P\left[ \chi ,\sigma |\chi
_{0},t_{0}\right] }{\delta \xi _{x}^{\bot }}\delta \xi _{x}^{\bot }\right)
\rho _{\sigma _{0}}\left[ \chi _{0}\right]
\end{eqnarray}%
which, using (\ref{Evol eq b}) and the fact that $\sigma ^{\prime }$ (or $%
\delta \xi _{x}^{\bot }$) can be freely chosen leads to the local
Fokker-Planck equations, 
\begin{equation}
\frac{\delta \rho _{\sigma }\left[ \chi \right] }{\delta \xi _{x}^{\bot }}=%
\frac{\eta }{g_{x}^{1/2}}\left\{ -\frac{\delta }{\delta \chi _{x}}\left(
\,\rho _{\sigma }\left[ \chi \right] \frac{\delta \phi \left[ \chi \right] }{%
\delta \chi _{x}}\right) +\frac{1}{2}\frac{\delta ^{2}}{\delta \chi _{x}^{2}}%
\rho _{\sigma }\left[ \chi \right] \right\} ~,  \label{FP 3}
\end{equation}%
which is equation (\ref{FP equation}).


\begin{thebibliography}{99}
\bibitem{Birrell Davies 1984} N. D. Birrell and P. C. W. Davies, \emph{%
Quantum Fields in Curved Space }(Cambridge U.P.,\ Cambridge 1984).

\bibitem{Wald 1994} R. M. Wald, \emph{Quantum Field Theory in Curved
Spacetime and Black Hole Thermodynamics} (U. Chicago Press, Chicago 1994).

\bibitem{Hollands Wald 2014} S. Hollands and R. M. Wald, \textquotedblleft
Quantum fields in curved spacetime,\textquotedblright\ in \emph{General
Relativity and Gravitation -- A Centennial Perspective} ed. by A. Ahtekar,
B. K. Berger, J. Isenberg, and M. MacCallum (Cambridge U.P., 2015); extended
online version: arXiv:1401.2026 [gr-qc].

\bibitem{Fulling 1973} S. A. Fulling, Phys. Rev. D \textbf{7}, 2850 (1973).

\bibitem{Unruh 1976} W. G. Unruh, Phys. Rev. D \textbf{14}, 870 (1976).

\bibitem{Bekenstein 1973} J. D. Bekenstein, Phys. Rev. D \textbf{7}, 2333
(1973).

\bibitem{Bardeen 1973} J. M. Bardeen, B. Carter, and S. W. Hawking, Comm.
Math. Phys. \textbf{31}, 161 (1973).

\bibitem{Hawking 1976} S. W. Hawking, Comm. Math. Phys. \textbf{43}, 199
(1975), erratum: \emph{ibid}. \textbf{46}, 206 (1976).

\bibitem{Schlosshauer 2004} M. Schl\"{o}sshauer, Rev. Mod. Phys. \textbf{76}%
, 1267 (2004).

\bibitem{Jaeger 2009} G. Jaeger, \emph{Entanglement, Information, and the
Interpretation of Quantum Mechanics} (Springer-Verlag, Berlin Heidelberg
2009).

\bibitem{Leifer 2014} M. S. Leifer, Quanta \textbf{3}, 67 (2014); arXiv.org:
1409.1570.

\bibitem{Nelson 1985} E. Nelson, \emph{Quantum Fluctuations }(Princeton UP,
Princeton, 1985).

\bibitem{Adler 2004} S. Adler, \emph{Quantum Theory as an Emergent Phenomenon%
} (Cambridge UP, Cambridge, 2004).

\bibitem{Smolin 2006} L. Smolin, \textquotedblleft Could quantum mechanics
be an approximation to another theory?\textquotedblright\
arXiv.org/abs/quant-ph/0609109.

\bibitem{de la Pena Cetto 2014} L. de la Pe\~{n}a, A.M. Cetto, \emph{The
Emerging Quantum: The Physics Behind Quantum Mechanics }(Springer, 2014).

\bibitem{Grossing 2008} G. Gr\"{o}ssing, Phys. Lett. A \textbf{372}, 4556
(2008), arxiv:0711.4954; G. Gr\"{o}ssing \emph{et al}, J. Phys. Conf. Ser. 
\textbf{361}, 012008 (2012).

\bibitem{EmQm 2015} See \emph{e.g.}, the proceedings of the conference on 
\emph{EmQm15: Emergent Quantum Mechanics} \emph{2015}, J. Phy. Conf. Ser. 
\textbf{701} (2016) -- http://iopscience.iop.org/issue/1742-6596/701/1.

\bibitem{tHooft 2016} G. 't Hooft, \emph{The Cellular Automaton
Interpretation of Quantum Mechanics} (Springer, 2016).

\bibitem{Wootters 1981} W. K. Wootters, Phys. Rev. D \textbf{23}, 357-362
(1981).

\bibitem{Caticha 1998} A. Caticha, Phys. Lett. A \textbf{244}, 13-17 (1998);
Phys. Rev. A \textbf{57}, 1572-1582 (1998); Found. Phys. \textbf{30},
227-251 (2000).

\bibitem{Brukner Zeilinger 2002} C. Brukner, A. Zeilinger, \textquotedblleft
Information and Fundamental Elements of the Structure of Quantum
Theory\textquotedblright\ in\ \emph{Time, Quantum, Information}, ed. L.
Castell, and O. Ischebeck, (Springer 2003); arXiv:quant-ph/0212084.

\bibitem{Hall Reginatto 2002} M.J.W. Hall, M. Reginatto, J. Phys. A \textbf{%
35}, 3289-3299 (2002); Fortschr. Phys. \textbf{50}, 646-656 (2002).

\bibitem{Spekkens 2007} R. Spekkens, Phys. Rev. A \textbf{75}, 032110 (2007).

\bibitem{Goyal Knuth Skilling 2010} P. Goyal, K. Knuth, and J. Skilling,
Phys. Rev. A \textbf{81}, 022109 (2010).

\bibitem{Chiribela et al 2011} G. Chiribella, G. M. D'Ariano, and P.
Perinotti, Phys. Rev. A\textbf{\ 84}, 012311 (2011).

\bibitem{Reginatto Hall 2011} M. Reginatto and M.J.W. Hall, AIP Conf. Proc. 
\textbf{1443}, 96 (2012); arXiv:1108.5601.

\bibitem{Reginatto Hall 2012} M. Reginatto and M.J.W. Hall, AIP Conf. Proc. 
\textbf{1553}, 246 (2013); arXiv:1207.6718.

\bibitem{Hardy 2013} L. Hardy, \textquotedblleft Reconstructing Quantum
Theory,\textquotedblright\ in \emph{Quantum Theory: Informational
Foundations and Foils} ed. G. Chiribella and R. Spekkens (Springer, 2015);\
arXiv:1303.1538.

\bibitem{DAriano 2017} G. M. D'Ariano, Int. J. Th. Phys. \textbf{56}, 97
(2017)

\bibitem{Caticha 2010} A. Caticha, J. Phys. A: Math. Theor. \textbf{44},
225303 (2011); arXiv.org:1005.2357.

\bibitem{Caticha 2015} A. Caticha, Entropy \textbf{17}, 6110 (2015);
arXiv.org:1509.03222.

\bibitem{Caticha 2017} A. Caticha, Annalen der Physik, (2018), 1700408.
https://doi.org/10.1002/andp.201700408; arXiv:1711.02538v2.

\bibitem{Jaynes 1957} E. T. Jaynes, Phys. Rev. \textbf{106}, 620 and \textbf{%
108}, 171 (1957);

\bibitem{Jaynes 1983} \emph{E. T. Jaynes: Papers on Probability, Statistics
and Statistical Physics}, Ed. by R. D. Rosenkrantz (Reidel, Dordrecht, 1983).

\bibitem{Caticha 2012} For a pedagogical introduction see A. Caticha, \emph{%
Entropic Inference and the Foundations of Physics} (EBEB 2012, S\~{a}o
Paulo, Brazil); http://www.albany.edu/physics/ACaticha-EIFP-book.pdf.

\bibitem{Nelson 1979} E. Nelson, \textquotedblleft Connection between
Brownian motion and quantum mechanics,\textquotedblright\ \emph{Einstein
Symposium Berlin}, Lect. Notes Phys. \textbf{100}, 168 (Springer-Verlag,
Berlin, 1979).

\bibitem{Bartolomeo et al 2014} A. Caticha, D. Bartolomeo, M. Reginatto, AIP
Conf. Proc. \textbf{1641}, 155 (2015); arXiv.org:1412.5629.

\bibitem{Caticha 2012b} A. Caticha, AIP Conf. Proc. \textbf{1553}, 176
(2013); arXiv.org/abs/1212.6946.

\bibitem{Ipek Caticha 2014} S. Ipek and A. Caticha, AIP Conf. Proc. \textbf{%
1641}, 155 (2015); arXiv.org:1412.5637.

\bibitem{Ipek et al 2017} S. Ipek, M. Abedi, and A. Caticha, AIP Conf. Proc. 
\textbf{1853}, 090002 (2017); arXiv:1803.06327.

\bibitem{Dirac 1951} P.A.M. Dirac, Can. J. Math. \textbf{3}, 1 (1951).

\bibitem{Dirac Lectures} P.A.M. Dirac, \emph{Lectures on Quantum Mechanics }%
(Dover, New York, 2001).

\bibitem{Teitelboim 1973a} C. Teitelboim, Ann. Phys. \textbf{79}, 542 (1973).

\bibitem{Teitelboim 1973b} C. Teitelboim, \textquotedblleft The Hamiltonian
structure of spacetime\textquotedblright\ Ph.D. thesis, Princeton University
(1973).

\bibitem{Kuchar 1973} K. Kucha\v{r}, \textquotedblleft Canonical
Quantization of Gravity\textquotedblright\ in \emph{Relativity,
Astrophysics, and Cosmology}, p. 237-288, W. Israel (ed.) (Reidel, Dordrecht
1973).

\bibitem{Hojman et al 1976} S. A. Hojman, K. Kucha\u{r}, and C. Teitelboim,
Ann. Phys. \textbf{96}, 88 (1976).

\bibitem{Weiss 1938} P. Weiss, Proc. Roy. Soc. London A\textbf{\ 169}, 107
(1938).

\bibitem{Tomonaga 1946} S. Tomonaga, Progr. Theor. Phys. \textbf{1}, 27
(1946).

\bibitem{Dirac 1948} P.A.M. Dirac, Phys. Rev. \textbf{73}, 1098 (1948).

\bibitem{Schwinger 1948} J. Schwinger, Phys. Rev. 1439 (1948). 

\bibitem{Johnson Caticha 2011} D.T. Johnson and A. Caticha, AIP Conf. Proc. 
\textbf{1443}, 104 (2012); arXiv:1108.2550

\bibitem{Vanslette Caticha 2016} K. Vanslette and A. Caticha, AIP Conf.
Proc. \textbf{1853}, 090003 (2017); arXiv:1701.00781.

\bibitem{Nawaz Caticha 2011} S. Nawaz and A. Caticha, AIP Conf. Proc. 
\textbf{1443}, 112 (2012); arXiv:1108.2629. 

\bibitem{Nawaz et al  2015} S. Nawaz, M. Abedi, and A. Caticha, AIP Conf.
Proc. \textbf{1757}, 030004 (2016); arXiv.org:1601.01708. 

\bibitem{Bartolomeo Caticha 2015} D. Bartolomeo and A. Caticha, AIP Conf.
Proc. \textbf{1757}, 030002 (2016); arXiv.org:1512.09084.

\bibitem{Bartolomeo Caticha 2016} D. Bartolomeo and A. Caticha, J. Phys:
Conf. Series \textbf{701}, 012009 (2016); arXiv.org:1603.08469. 

\bibitem{Demme Caticha 2016} A. Demme and A. Caticha, AIP Conf. Proc. 
\textbf{1853}, 090001 (2017); arXiv.org:1612.01905. 

\bibitem{Caticha Carrara 2019} A. Caticha and N. Carrara, \textquotedblleft
The Entropic Dynamics of Spin,\textquotedblright\ in preparation (2019).

\bibitem{Wald 1984} R. M. Wald, \emph{General Relativity} (U. Chicago Press,
Chicago, 1984).

\bibitem{Ipek 2019} S. Ipek, \textquotedblleft \emph{The Entropic Dynamics
of Relativistic Quantum Fields in Curved spacetime,}\textquotedblright\
(Ph.D. thesis, University at Albany 2019), work in progress. 

\bibitem{Hall et al 2003} M.J.W. Hall, K. Kailash, and M. Reginatto,\ J.
Phys. A: Math. Gen. \textbf{36}, 9779 (2003); arXiv.org:hep-th/0307259.

\bibitem{Long Shore 1998} D. V. Long and G. M. Shore, Nucl. Phys. B\textbf{\
530}, 247 (1998).

\bibitem{Gourgoulhon 2007} E. Gourgoulhon, \textquotedblleft 3+1 formulation
and basis of numerical relativity,\textquotedblright\
arXiv.org:gr-qc/0703035 (2007).

\bibitem{Jackiw 1989} R. Jackiw, \textquotedblleft Analysis on
infinite-dimensional manifolds --- Schr\"{o}dinger representation for
quantized fields\textquotedblright\ in \emph{Field Theory and Particle
Physics}, 5th Jorge Swieca Summer School, Brazil 1989, ed. by O. \'{E}boli 
\emph{et al}. (World Scientific, 1990).

\bibitem{Freese et al 1985} Freese, Katherine, Christopher T. Hill, and Mark
Mueller. ``Covariant functional Schr\~{A}\P dinger formalism and application
to the Hawking effect." Nuclear Physics B 255 (1985): 693-716.

\bibitem{Guven et al 1989} Guven, Jemal, Bennett Lieberman, and Christopher
T. Hill. ``Schr\~{A}\P dinger-picture field theory in Robertson-Walker flat
spacetimes." Physical Review D 39.2 (1989): 438.

\bibitem{Floreanini et al 1987} Floreanini, R., Chris T. Hill, and R.
Jackiw. ``Functional representation for the isometries of de Sitter space."
Annals of Physics 175.2 (1987): 345-365.

\bibitem{Halliwell 1991} Halliwell, Jonathan J. ``Global spacetime
symmetries in the functional Schr\~{A}\P dinger picture." Physical Review D
43.8 (1991): 2590.

\bibitem{Ballentine 1998} Ballentine, Leslie E. \emph{Quantum mechanics: a
modern development.} (World Scientific Publishing Company, 1998).

\bibitem{Ashtekar Schilling 1999} A. Ashtekar and T. A. Schilling,
\textquotedblleft Geometrical Formulation of Quantum
Mechanics,\textquotedblright\ in \emph{On Einstein's Path}, p. 23-65, A.
Harvey (ed.) (Springer, New York 1999).

\bibitem{Callender Huggett 2001} Huggett, Nick, and Craig Callender. ``Why
quantize gravity (or any other field for that matter)?." Philosophy of
Science 68.S3 (2001): S382-S394.

\bibitem{Mattingly 2005} Mattingly, James. ``Is quantum gravity necessary?."
The universe of general relativity. Birkh\"{a}user Boston, 2005. 327-338.

\bibitem{Wuthrich 2005} W\"{u}thrich, Christian. ``To quantize or not to
quantize: fact and folklore in quantum gravity." Philosophy of Science 72.5
(2005): 777-788.

\bibitem{Carlip 2008} Carlip, Steve. ``Is quantum gravity necessary?."
Classical and Quantum Gravity 25.15 (2008): 154010.

\bibitem{Albers et al 2008} Albers, Mark, Claus Kiefer, and Marcel
Reginatto. ``Measurement analysis and quantum gravity." Physical Review D
78.6 (2008): 064051. arXiv: 0802.1978.

\bibitem{Kibble 1979} T. Kibble,\ Comm. Math. Phys. \textbf{65}, 189 (1979).

\bibitem{Heslot 1985} A. Heslot, Phys. Rev. D\textbf{\ 31}, 1341 (1985).

\bibitem{Ipek Caticha 2019} S. Ipek and A. Caticha, \textquotedblleft The
Entropic Dynamics approach to Semi-Classical Gravity,\textquotedblright\ in
preparation (2019).
\end{thebibliography}
\end{document}